# Quantum Games and Interactive Tools for Quantum Technologies Outreach and Education

Zeki C. Seskir,[a,*] Piotr Migdał,[b] Carrie Weidner,[c] Aditya Anupam,[d] Nicky Case,[e] Noah Davis,[f] Chiara Decaroli,[g] İlke Ercan,[h] Caterina Foti,[i,j] Paweł Gora,[k] Klementyna Jankiewicz,[b] Brian R. La Cour,[f] Jorge Yago Malo,[l] Sabrina Maniscalco,[i,j,m] Azad Naeemi,[n] Laurentiu Nita,[o] Nassim Parvin,[d] Fabio Scafirimuto,[p] Jacob F. Sherson,[q,r] Elif Surer,[s] James Wootton,[p] Lia Yeh,[t,u] Olga Zabello,[v] Marilù Chiofalo[i,l]

[a] Karlsruhe Institute of Technology, ITAS, Karlstraße 11, Karlsruhe, Germany, 76133
[b] Quantum Flytrap, ul. Ceglowska 29, Warsaw, Poland, 01-809
[c] Quantum Engineering Technology Laboratories, H.H. Wills Physics Laboratory and Department of Electrical and Electronic Engineering, University of Bristol, Bristol, United Kingdom, BS8 1FD
[d] Georgia Institute of Technology, School of Literature, Media, and Communication, 686 Cherry St NW, Atlanta, GA, United States, 30313
[e] *https://ncase.me/*
[f] The University of Texas at Austin, Center for Quantum Research, Applied Research Laboratories, 10000 Burnet Rd, Austin, Texas, United States, 78758
[g] National Quantum Computing Center, Didcot, Oxfordshire, UK, OX11 0GD
[h] TU Delft, Department of Microelectronics, Mekelweg 4, Delft, Netherlands, 2628 CD
[i] QPlayLearn
[j] Aalto University, InstituteQ – The Finnish Quantum Institute, Department of Applied Physics, FI-00076, Aalto, Finland, 11000
[k] Quantum AI Foundation, Sanocka 9/103, Warsaw, Poland, 02-110
[l] University of Pisa, Department of Physics, Largo Bruno Pontecorvo 3, Pisa, Italy, I-56126
[m] University of Helsinki, InstituteQ – The Finnish Quantum Institute, Department of Physics, Helsinki, Finland
[n] Georgia Institute of Technology, School of Electrical and Computer Engineering, 777 Atlantic Drive NW, Atlanta, GA, United States 30332
[o] Quarks Interactive, Strada Bailor 93a, Miercurea-Ciuc, Harghita, Romania, 4500
[p] IBM Quantum, IBM Research - Zurich, Säumerstrasse 4, Rüschlikon, Switzerland, 8803
[q] ScienceAtHome, Fuglesangs Allé 4, Aarhus, Denmark, 8210
[r] Aarhus University, Institute for Physics and Astronomy, Ny Munkegade 120, Aarhus, Denmark, 8000
[s] Middle East Technical University, Graduate School of Informatics, Department of Modeling and Simulation, Üniversiteler Mah. Dumlupınar Blv. No:1, Ankara, Turkey, 06800
[t] University of Oxford, Department of Computer Science, Wolfson Building, Parks Road, Oxford, UK, OX1 3QD
[u] IBM Quantum, IBM T. J. Watson Research Center, Yorktown Heights, NY, USA, 10598
[v] Offenburg University of Applied Sciences, Department of Business and Industrial Engineering, Badstraße 24, Offenburg, Germany, 77652

*Zeki C. Seskir, E-mail: zeki.seskir@kit.edu

**Abstract.** In this article, we provide an extensive overview of a wide range of quantum games and interactive tools that have been employed by the community in recent years. The paper presents selected tools, as described by their developers. The list includes *Hello Quantum, Hello Qiskit, Particle in a Box, Psi and Delta, QPlayLearn, Virtual Lab by Quantum Flytrap, Quantum Odyssey, ScienceAtHome,* and *The Virtual Quantum Optics Laboratory*. Additionally, we present events for quantum game development: hackathons, game jams, and semester projects. Furthermore, we discuss the Quantum Technologies Education for Everyone (QUTE4E) pilot project, which illustrates an effective integration of these interactive tools with quantum outreach and education activities. Finally, we aim at providing guidelines for incorporating quantum games and interactive tools in pedagogic materials to make quantum technologies more accessible for a wider population.

## 1 Introduction

The second quantum revolution[1] has been unfolding for over two decades, and numerous countries across the globe acknowledge the potential of these technologies. In the last decade, more than fifteen publicly supported national initiatives were launched with several of them having budgets exceeding billion euros, particularly in the UK[2], the US[3], the EU[4], France[5], Germany[6], India[7], and China[8]. Both the number of new publications per year[9] and patents[10] granted on quantum technologies (QT) have also been climbing steadily for the last decade, along with the growing number of start-ups and renowned companies getting into the field. Recently, the field has been receiving attention from private funding sources, which some have been referring to as the 'quantum gold rush'[11]. Within two decades, QT has advanced from a topic that was mainly discussed in physics conferences to become a strong contender for a potential future industry.

The developments in QT, like most emerging technologies, bring along several social and technical challenges. First and foremost, for the general public, QT is still an abstract concept with limited accessible material on the topic[12]. However, there is an urgency to make QT accessible to lay audiences and facilitate the stakeholders' inclusion into the debate on the subject[13]. Secondly, the growing advancements in the industry require a rapid expansion of the workforce[14] and an increased level of understanding of QT in educational, business, and policy-making efforts. Lastly, like most promising emerging technologies, QT is expected to impact society profoundly[15]; however, this may also cause preexisting divides on national and international levels to grow wider. These concerns encourage some members of the quantum community to adopt a responsible approach[16] and build a common language[17].

One of the most important steps taken towards addressing these concerns is to increase public engagement and involvement in QT. The predominant ideas in the existing literature are on increasing quantum literacy at the K-12 and general public[18] levels, and focusing more on education and developing transdisciplinary problem-solving skills[19]. Therefore, education and outreach activities are of paramount importance in promoting QT and making it accessible for a wider audience, engaging all potential stakeholders.

The literature on QT and its 'social' impact has expanded significantly[20], and stakeholders' engagement in QT is rapidly increasing, therefore numerous initiatives are being set forth to address these growing demands. There are several coordinated efforts around the world, such as the quantum technologies education coordination and support action for the EU's Quantum Flagship (QTEdu CSA)[21], the Q-12 Education Partnership in the US[22], and the IEEE Quantum Education[23] initiative. In addition to organizing extracurricular activities, these initiatives also focus on identifying and improving the potential tools that can be utilized by quantum outreach and education practitioners. In this respect, learning quantum concepts through games has gained prominence as an approach to benefit the quantum community at large and support these outreach and education efforts extensively. For example, quantum games are a key means of engaging students and teachers, making up four out of five of the Q-12 Education Partnership's QuanTime K-12 classroom activities[24]. These quantum games are used here to refer to computer (or video) games with one or more of the phenomena from quantum physics embedded in their game mechanics —note that this is not (in principle) related to research efforts extending game theory to include quantum mechanics[25,26].

The development of science-based games can set out to achieve a variety of goals, including for educational purposes, citizen-science-based[27] research purposes, science outreach and public engagement, and/or fun. Games are unique among other tools for research and education in that, if they are widely adopted, they have the potential to reach audiences beyond most coordination efforts, and thus they can be useful vectors for increasing quantum awareness. This is important for two reasons. First, it is beneficial that the goal of understanding the science target not only current and future technical experts on the topic, but also people with other occupations and skills who will transform the science into technologies impacting many other areas. Equally important is for the learner to gain exposure to a new concept, thereby making the concepts less intimidating when reencountered. This is especially impactful to learners from underrepresented or disadvantaged backgrounds who had less opportunities or encouragement to experience this topic in the first place, for them to succeed in the field utilizing their skills and perspectives.

In this article, we introduce a wide range of tools that have been tested by developers and used by the community for several years. We present different approaches and tools, together with the experiences of their developers, aiming to provide current and future practitioners of quantum outreach and education activities a better understanding of their application and extent. Most of the authors are members of the pilot project *Quantum Technologies Education for Everyone*[28] (QUTE4E) organized under the Quantum Technology Education Coordination and Support Action (QTEdu CSA) of the EU's Quantum Flagship programme. Therefore, the list of games and interactive tools provided here consists of a non-exhaustive, but representative list. Further information on the pilot project can be found in Section 5.

The article is formatted in the following manner. In Section 2, we argue for the importance of photonics in QT outreach and education efforts. In Section 3, we discuss the relationship between science-based games and quantum games[29], including a brief discussion on the evaluation of quantum games. In Section 4, subsections 4.1-4.7 are dedicated to introducing quantum games and interactive tools for QT outreach and education, and in subsection 4.8 we explore some activities in creating quantum games. Finally, in Section 5, a more detailed description of the pilot project QUTE4E is given.

## 2 Locating Photonics in QT Outreach and Education

To understand quantum technologies—such as quantum cryptography, quantum computing or quantum sensing— a knowledge of the quantum phenomena they rely on is essential. These relevant quantum concepts include superposition, entanglement, interference, dynamical evolution and measurement[30–33], and there is an emerging literature introducing these through optics concepts[34–38].

Using qubits allows us to discuss quantum information without relating it to physics[39]. While this adds simplicity in some respects, it also results in challenges. First, all current quantum devices are deeply rooted in physics. For now, to use them efficiently, we need to consider which operations are physically possible. Second, quite a few phenomena are easier to explain with concrete systems. This offers a direct way to visualize them and relate them to the broader fields of wave physics, electrodynamics, and optical communications.

Introductions that aim to be entry-level usually use electron spin[40] or photon polarization[41] to introduce quantum mechanics. Arguably[42] photon polarization is an easier two-level system that translates to demonstrable experiments with classical light and behaves more intuitively with rotations and superposition.

Quantum mechanics is known to be prone to misconceptions, resulting from otherwise well-intentioned physicists, teachers and science communicators[43]. Misconceptions are common even among students of physics[44,45]. A common problem with teaching quantum mechanics lies in training students to distinguish the phase of a superposition. Telling someone that it is possible to have a qubit in the $|0\rangle$ and $|1\rangle$ state simultaneously leaves no such space and fails to distinguish between statistical uncertainty and quantum mechanics, as explained by Scott Aaronson[46].

> *Over the years, I've developed what I call the Minus-Sign Test, a reliable way to rate popularizations of quantum mechanics. To pass the Minus-Sign Test, all a popularization needs to do is mention the minus signs: i.e., interference between positive and negative amplitudes, the defining feature of quantum mechanics, the thing that makes it different from classical probability theory, the reason why we can't say Schrödinger's cat is "really either dead or alive," and we simply don't know which one, the reason why the entangled particles can't have just agreed in advance that one would spin up and the other would spin down.*

However, photon polarization gives a quick way to distinguish such states and to show that they are orthogonal (see Figure 1).

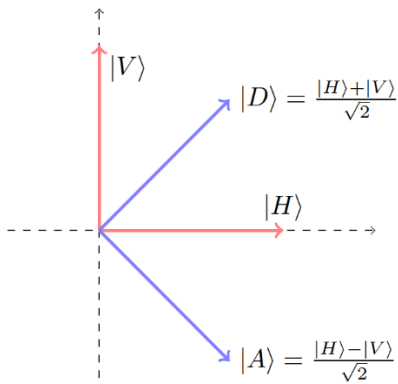

Figure 1. Polarization of light offers a straightforward way to present orthogonal states. With a typical mapping $|H\rangle=|0\rangle$ and $|V\rangle=|1\rangle$, quantum states $(|0\rangle\pm|1\rangle)/\sqrt{2}$ have a direct physical representation, both easily demonstrable experimentally in a class and matching the high-school definition of orthogonality.

Moreover, such systems give an easy way to explain measurements (with linear polarization filters known in photography and LCD screens, followed by detectors known from cameras), translate back and forth with non-local superposition over different paths (via polarizing beam splitters), and relate to simpler and laboratory-demonstrable experiments with interference. Additionally, several quantum communication protocols were first demonstrated with photons. These include quantum key distribution with BB84[47] and Ekert[48] protocols, the Bell inequality violation[49,50], and boson sampling[51].

Photonic systems can be easily extended to more advanced systems. For example, to encode qudits (quantum systems with d levels, typically d>2), one uses different spatial modes and angular momentum, giving rise to qualitatively different classes of entanglement than that achievable with qubits[52]. Photonics can also be used to simulate molecular vibrations and dynamics[53].

At the same time, it is crucial to underline that there are many other physical implementations of quantum technologies - involving spin, energy levels, and currents in superconducting loops, among others. We believe that to teach modern quantum technologies we need to show phenomena from different angles — both physical and abstract. The same core properties of quantum mechanics are important no matter which system we study, yet some may be better to explain a given phenomenon, or more easily relatable to currently available quantum devices and protocols. Hence, we present a wide array of quantum games and quantum interactive tools, each approaching from various angles, underlying various phenomena.

It should also be noted that there is an emerging literature on introducing quantum technologies (especially quantum computation) through computer science[54–56], programming[57–60] oriented, and visually-oriented (no-math) approaches[19,61–63] in parallel to the physics-focused approaches. All of which can be thought of as efforts in further developing the outreach and education landscape of quantum technologies.

# 3  Quantum Games and Interactive Tools as Novel Methods for QT Outreach and Education

We believe that learning the basic concepts of quantum physics and quantum information science should be made accessible such that the public can understand the significance of the second quantum revolution[13], and anyone regardless of their background can access the tools to learn and develop quantum technologies. There are many learning resources available[64], but to attain understanding, it is crucial that the learning be sequential[65]—someone learning quantum computing for the first time would not know that to understand quantum algorithms, one first has to learn about qubits, then quantum gates, and so on. Games are a natural way for pacing progression according to how well the learner grasps these concepts. Moreover, there need to be efforts made in establishing channels of dissemination for these learning opportunities to reach a broader audience. Games also support this goal due to their horizontal reach in society.

The interdisciplinary nature of the QT field indicates that the quantum workforce calls for a vastly diverse range of skills[14], and so every effort should be made to include people from different backgrounds. This poses a particular challenge: teaching quantum mechanics is daunting to teachers for whom this is also a new topic. Virtual labs and learning tools, accompanied by explanations of the physics being simulated, are a cost-effective means of providing an environment for exploring and building intuition of quantum phenomena.

The connection between interactive storytelling and quantum computing was briefly explored in the literature[66], however, a comprehensive discussion on why and how these novel tools can be utilized for education and outreach purposes has not been made. We would like to provide the groundwork for such a discussion, by providing the necessary background, with a quick introduction to science-based (Subsection 3.2) and serious games (Section 4), followed by description of quantum games and interactive tools from their developers (Subsections 4.1-4.7).

In this section, we cover four topics. First, we discuss some early and current quantum tools in subsection 3.1. The early tools can be accepted as precursors to the quantum games and interactive tools of today. Second, we briefly discuss the wider arena of science-based games in subsection 3.2. It is essential to realize that this shift toward using games for outreach and education purposes for science topics is not limited to quantum information science but indeed a general trend. The increase in the number of science-based games and the comparative success of some of these games are valuable signifiers of this trend. In subsection 3.3, we provide *Explorable Explanations* as an example of how added interactivity can play a practical role in science communication. Finally, in subsection 3.4 we discuss the challenge of evaluating quantum games and provide some suggestions for future work to be done in this arena.

## 3.1 Interactive tools

A number of different computer-based quantum tools have been developed for educational purposes. Typically, these tools are geared towards the visualization of quantum phenomena. Early efforts by the Consortium for Upper Level Physics Software released quantum visualization software packages in the mid-1990s[67,68]; some of the authors of these software packages also released a book on quantum mechanics simulation[67]. Later efforts include a series of books by Thaller[69,70] and Zollman[71].

In the early 2000s, Paul Falstad released a series of Java applets simulating physics[72], including 12 on quantum mechanics and many others on waves physics. In particular, the 1-D Quantum Mechanics Applet allows one to draw any potential, any wavefunction, and explore the position space, momentum space, energies and eigenstates simultaneously. The Ripple Tank (2-D Waves) Applet allows exploration of interference —crucial for both classical wave physics and quantum phenomena. Unlike other simulations, a user can draw the interference setup in a sandbox mode. Paul Falstad's applets gained enough popularity to be ported as mobile apps and rewritten in JavaScript so as to be usable with modern browsers.

Other web-based applications include the Physlet simulations[73] and a set of simulations developed by Joffre[74]. The ubiquitous PhET simulations[75] have developed a set of quantum-specific simulations; PhET researchers have published a wide array of research on the design and use of their simulation[76–79]. Other such research-validated quantum simulation tools include the QuILTs[80] developed at the University of Pittsburgh[81], the QuViS tutorials[82] developed at St. Andrew's University[83,84], and the Quantum Composer simulation tool[85] developed at Aarhus University[86,87], which is described in more detail later in this paper.

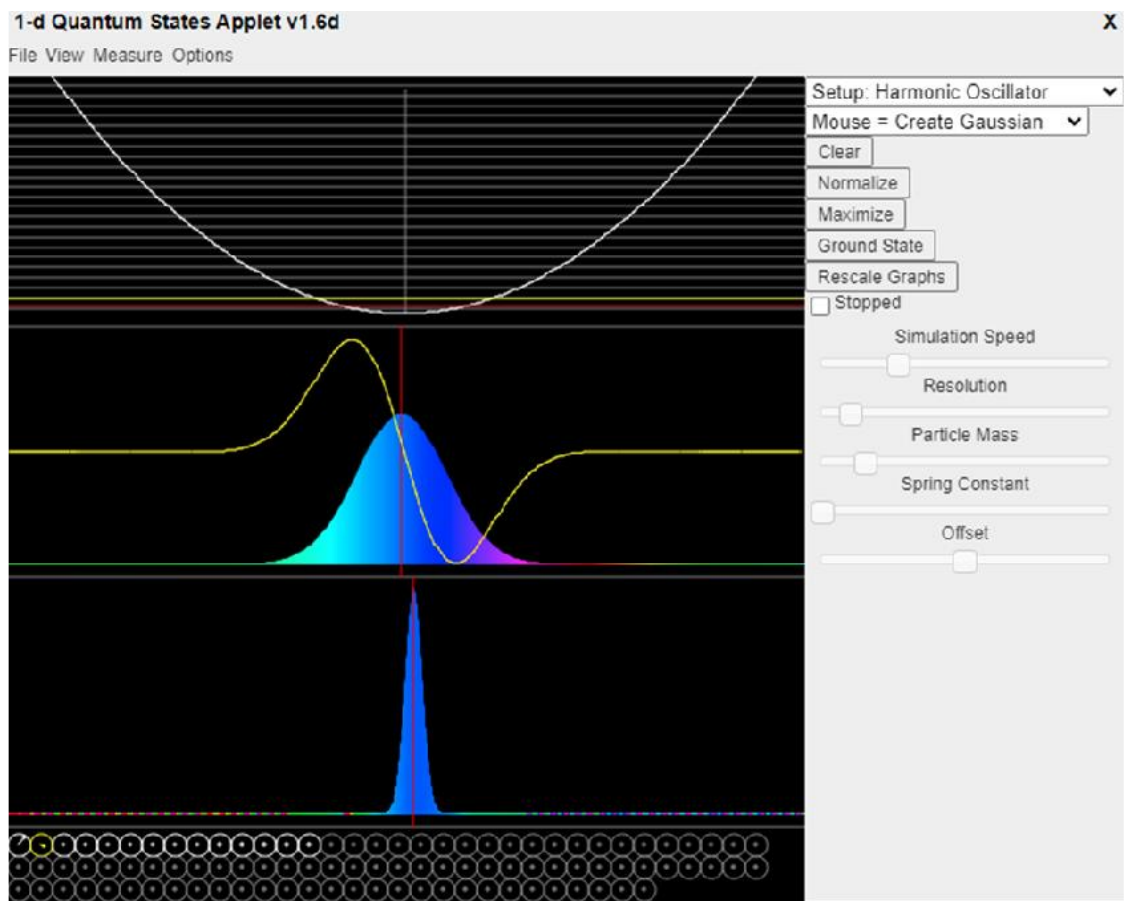

Figure 2. 1-D Quantum Mechanics Applet (a single-particle quantum mechanics states in one dimension) by Paul Falstad (2002), available at https://www.falstad.com/qm1d/. An early simulation gives a user the ability to simultaneously see (and edit): one-dimensional potential, eigenstates with eigenenergies, wavefunction as a function of position and time, wavefunction as a function of momentum, relative amplitudes of each eigenstate.

With the exception of the node-based programming in Quantum Composer, the tools developed here are typically set modules exploring a specific quantum phenomenon (sometimes including supplementary material that instructors can use to develop exercises), and interactive tools (e.g. the PhET, QuViS, and QuILT simulations) allow students to easily modify parameters and explore the simulation in more detail. However, none of these tools can be classified as games, as they lack the required elements in the interface and design that engage the students outside of their traditional educational goals. In Section 4, we explore more recent and ongoing efforts in accomplishing this task, but first, we introduce the concept of science-based games and *Explorable Explanations* as an example.

## 3.2 Science-based games

The traditional way of teaching includes textbooks, lectures, and projects. This teaching method is performed top-down, with clear instructions and expectations. Students are tested via grading of projects, homework assignments and exams. Usually fun, or the ability to explore on one's own, is not the key focus. Often students' motivation is extrinsic—to pass a course with a good grade. These activities can be gamified, provided with supplementary goals, scores, and challenges to make them more engaging.

In contrast, science-based games[29] approach teaching from the opposite direction: to primarily focus on creating an experience sparking intrinsic motivation[88]. That is, students play for fun, but learn in the process, as their gaming experience requires learning concepts to proceed or provides an explorative pathway through the game that promotes learning[89] (even if it is not strictly necessary).

There are several existing games showing different aspects of physics—special relativity theory[90], electromagnetism[91], classical physics and orbital mechanics[92]. There are games that teach computing - creating circuits from the simplest blocks - NAND gates[93] and low-level programming in assembly. While a lot of games are standalone one-off projects, some game developers specialize in science-based games, e.g., Zachtronics (algorithmics) and Test Tube Games (physics). Some science-based games go way beyond the niche - SpaceChem and Kerbal Space Program got over 1 million and over 2 million sold copies on Steam[94], respectively.

## 3.3 Explorable Explanations

*Explorable Explanations*[95] (or "*explorables*") come at educational games from the opposite direction: instead of "games, but with science communication added", they're "science communication, but with interactivity added", although we should note that interactivity is a necessary but not sufficient condition for something to be considered a game. Most *explorables* are educational articles, with embedded simulations instead of static media such as images and videos. The term "*Explorable Explanation*" was coined in 2011 by interaction designer Bret Victor[96], with an ambitious goal of creating a two-way communication between the author and the reader. With an open-ended creative environment, readers can challenge the author's assumptions by going beyond what the author ever imagined. This ambition is explained more in the 2019 essay[97] by former Khan Academy designer Andy Matuschak and quantum physicist Michael Nielsen. These approaches are tightly related to exploring new visual languages of communication (see Figure 3).

$$X_k = \frac{1}{N} \sum_{n=0}^{N-1} x_n e^{i 2\pi k \frac{n}{N}}$$

To find the energy at a particular frequency, spin your signal around a circle at that frequency, and average a bunch of points along that path.

Figure 3. A self-descriptive formula of the discrete Fourier transform - a web-based MathJax implementation[98] of an equation color-coding scheme designed by Stuart Riffle from 2011[99]. Earlier approaches of color-coding go back to the Byrne's Euclid from 1847[100] subtitled *"In which Coloured Diagrams and Symbols are Used Instead of Letters for the Greater Ease of Learners"*.

Deep learning shares many similar concepts with QT - both hyped, fast-growing technologies, are heavily based on linear algebra with tensor products and have a non-trivial entry barrier. Abundance of online tools helping to learn and directly use deep learning already gets traction in gaming[101] and can provide guidance on how to present quantum technologies in an accessible, ready-to-use way. There are already numerous *Explorable Explanation*s in machine learning & AI[102], including general introductions to probability and statistics[103], decision trees and validation in machine learning[104], training of artificial neural network classifiers[105], and interpretability of convolutional neural networks[106,107]. In 2016-2021 a dedicated peer-reviewed journal *Distill* aimed at providing interactive explanations of novel research[108].

While the interactive environment of Jupyter Notebooks is a standard way of providing introductions to quantum software frameworks[109,110], there is only a handful of other quantum *Explorable Explanations*, e.g. a spaced-repetition based introduction to quantum computing[111], an introduction to quantum Fourier transform[112], and an exploration of single-qubit gates[113].

### 3.4 Evaluation of Quantum Games

A common question that one asks when considering any tool is whether it truly works as intended, and quantum games are no exceptions. Currently, there is no common design methodology[114] or an universal evaluation tool for quantum games, and this is largely due to the fragmentation in the field[115] (i.e., some games are developed professionally, some developed by university researchers, and some developed at hackathons, among others) and the differing goals between games (i.e., research, education, outreach, and/or fun). Indeed, some of the games described in Section 4 have been evaluated individually. For example, the developers of *Quantum Moves 2*, with its focus on citizen science, published their results in a peer-reviewed journal[116]. Similarly, the games *Particle in a Box*[61] and the *Quantum Odyssey*[117] have articles published in peer-reviewed journals containing evaluation sections. However, a proper and comprehensive evaluation and assessment scheme that can be applied to a variety of quantum games is missing in the literature. Validation of these quantum games, especially when ensuring that educational games serve to teach students relevant concepts, is important and should be further developed as activities in the field are progressing. While the development and discussion of a universal quantum game evaluation tool is beyond the scope of this paper, we offer some suggestions here.

First, we need to categorize quantum games in terms of their basic information such as on which platform they run on and their licenses, their types, and the concepts that they are drawing upon. Efforts to do this have been started[115], but the task remains difficult, as the landscape of quantum games is growing at an impressive rate, and games that are not actively developed upon quickly become obsolete (e.g., due to changes in operating system compatibility and security) or can be filled with game-breaking bugs. However, the development of a quantum games version of the Physport[118], which is a repository of research-based recommendations, assessments, tools, and practices for physics education, could be useful, and, as with the Physport, games can be given various ratings based on whether or not they have been evaluated (e.g., the Force Concept Inventory[119] is an assessment tool that has, due to its extensive research-based evaluation, received a gold star rating from Physport).

In parallel, researchers should work to develop best practices for the evaluation of quantum games (like those developed for quantum software[120]), focusing on the common goals of research, education, and fun; these best practices can be collected in the same repository as the games themselves. Finally, within the repository, we can also denote whether the game is being actively developed, is in a usable state, or has been deprecated in some way. This can be an invaluable tool to track how quantum games, their goals, and their use cases change over time.

Finally, evaluating the effectiveness and efficiency of a videogame for education requires a number of steps[121], which as a pilot we are completing. To start with, we need to work out a storytelling format, where the use of the purpose of the quantum game or interactive tool is made explicit. This has been the subject of theoretical work introducing the concept of culturo-scientific storytelling with special attention to quantum science and technologies[122], which is being tested in outreach events such as the workshops devoted to high-school students around the *Quantum Jungle* interactive installation[121]. Then, suitable evaluation tools must be designed in correspondence of on-purpose didactic experiments. For example, in one such experiment performed in a high school in Treviso (Italy), the *Quantum TiqTaqToe* videogame has been used as a central tool in a highly compact didactic intervention lasting just two afternoons (including a *TiqTaqToe* tournament), to educate students to the concepts of quantum state and measurement, quantum superposition and entanglement[123]. Suitable evaluation tools in the form of questionnaires have been designed for the purpose and analyzed, which show a high degree of effectiveness of the quantum game in supporting the students to develop an operational understanding of the aforementioned concepts[123]. While much systematic research needs to be performed along similar directions, these first steps appear quite promising and encouraging.

A preliminary effort to categorize the games presented in our paper is provided in Table 1.

| Title | Initial Release | Platform | License | Type | Key Concept(s) |
|---|---|---|---|---|---|
| Alice Challenge | 2017 | Web | Proprietary, Free | Citizen science game | Ultracold atoms, experimental optimization |
| Hello Qiskit / Hello | 2018 | Web / iOS / Android | Apache (web), Proprietary, | Puzzle game | Quantum gates |

| Quantum | | | Free (mobile) | | |
|---|---|---|---|---|---|
| Particle in a Box | 2016 | Web / MacOS / Windows | Proprietary, Free | Single player platformer game | Superposition, energy levels |
| Psi and Delta | 2018 | Web / MacOS / Windows | Proprietary, Free | Multiplayer platformer game | Superposition, energy levels, color-energy relationship |
| Quantum Game with Photons | 2014 | Web | MIT | Puzzle game | Photons |
| Quantum Odyssey | 2020 | Windows | Proprietary, Paid | Puzzle game | Quantum circuits |
| Quantum Moves 2 | 2018 | Web | Proprietary, Free | Citizen science game | Ultracold atoms, optimal control, wavefunctions and transport |
| Virtual Lab by Quantum Flytrap | 2019 | Web | Proprietary, Free | Simulator / Puzzle Game | Photons, Quantum Information, Measurement |

Table 1: A summary of games presented in Section 4.

# 4 Experiences From the Field of Quantum Games and Interactive Tools Development

Computer-based games have evolved from a niche pleasure for a select few to a worldwide phenomenon that is now a part of the everyday lives of millions of people. Games have surpassed movies and television shows as a cultural, sociological, and economic force to be reckoned with. Games represent a new, robust economic sector that has amassed significant money, technological know-how, and skill for many nations. Games User Research[124] has become an essential aspect of game production, and many of the industry's most notable companies now adopt this methodology. This translates to higher product quality, tremendous financial success, and a more positive work atmosphere, where employees can see evidence that their efforts are making a difference.

Serious games[125], an umbrella term encompassing games having additional aims beyond entertainment, are extensively employed in various fields, including healthcare, telerehabilitation, military, and education. Serious games have always focused on the physiological and spatial consequences of the games on the participants[126]. Serious games and game technologies have been utilized for simulation and training in various areas, and these technologies have been adapted to computer, virtual reality (VR) and mixed reality (MR) settings. Milgram et al.[127] offered a precise method for differentiating between various realities, and MR has been characterized as "mixed virtuality and augmented reality," as well as "any combination of real-world and virtual-world features." The application of gaming technology in VR and MR has enhanced spatial task research while also allowing immersion, presence, and usability studies.

The use of games and serious games has been a recent and powerful tool for experimenting with new quantum algorithms, providing new communication protocols, and teaching quantum theory to the masses. Recent advancements in implementing and testing quantum games have been an ongoing effort. In the rest of this section, we provide a selection of such games described by their developers.

## 4.1 Hello Quantum / Hello Qiskit

The *IBM Quantum Composer* was released in 2016, providing a simple and graphical way to write quantum circuits and run them on prototype quantum computers. This service was (and is) available for free to the general public, allowing them unprecedented access to quantum computer hardware[128]. However, without the knowledge of what quantum circuits were or how quantum gates worked, there was still a significant barrier to entry.

The *Hello Quantum* project, a collaboration between IBM Research and the University of Basel, was intended to alleviate this issue. The idea was to visualize quantum states and the effects of gates, so that people could gain an intuitive understanding of simple quantum circuits. This interactive and visual method for learning would be presented as a puzzle game, with the player being challenged to find the circuit that would take an input state to a desired output state.

The resulting products were released in four stages:
- A simple ASCII prototype, used in lectures at the University of Basel in 2017;
- The app *Hello Quantum*, released on iOS and Android in 2018[129] (Figure 4);
- The section *Hello Qiskit*, released in 2019 as part of the interactive Qiskit textbook[130];
- The section `Visualizing Entanglement', released in 2021 as part of the interactive Qiskit introduction course.

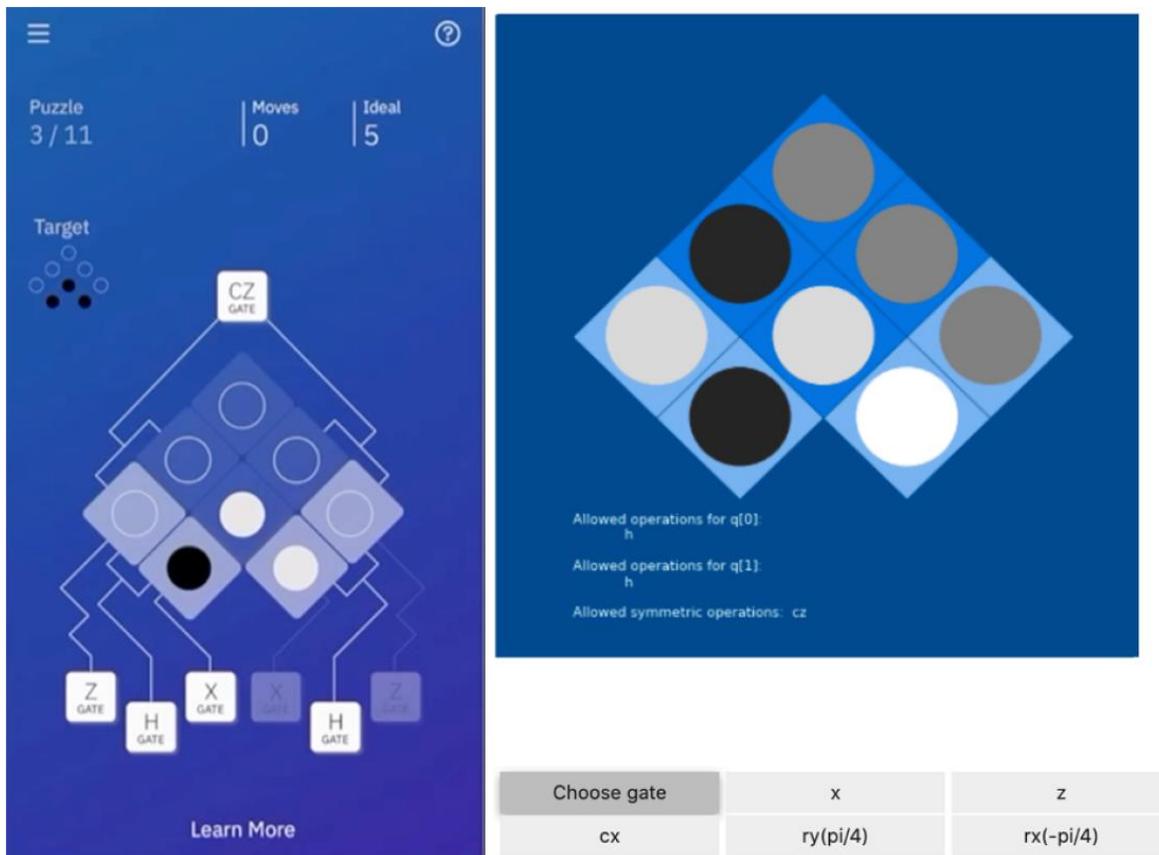

Figure 4. Screenshot of the Hello Quantum app (left) and a Jupyter widget-based puzzle from the Hello Qiskit section of the Qiskit textbook (right).

In each case, the learning material was presented by providing both interactive puzzles, intended to build intuition and provide context, as well as text providing a text explanation of how aspects of the game relate to quantum computing.

For `Hello Quantum', the puzzles and text were delivered entirely separately. The idea was that the player could play through the puzzles as a short and informal game. Then, once they had mastered the game mechanic, interested players could go on to read about how their new in-game skills translate to quantum computation. This text was not limited to the app itself, but instead represented the inaugural posts of the Qiskit blog. These posts were entitled "Visualizing bits and qubits"[131], which explained the visualization used in the app, and "Getting started with the IBM Q Experience"[132], which explained how to transfer the learned skills into what is now called the IBM Quantum Composer.

It was expected that the readers would be led to the former blog post by the app, and to the latter by a link in the former. However, view statistics show that only around 1% of views for the former can be traced to following the in-app link. Instead, the majority (around 58%) came from search engines. For the latter post, 81% of views come from sources external to the blog, such as 40% through email, instant message (IM) and direct and around 27% from search engines.

Given these results, we can conclude that the blog posts were a more successful learning resource than the app. This is since every app user who went on to fully understand the game will have followed the link to the app, but these app users are only a small percentage of the total readership.

For the 'Hello Qiskit' section of the Qiskit textbook, the puzzles were implemented within an interactive Jupyter notebook. This was located in a chapter specifically for interactive demos. An analysis of views for the textbook was conducted for the period March-November 2020[60]. It was found that the view share of this section is very comparable to those of the first two chapters. In fact, the average number of views for sections in chapters 1 and 2 during the period is very similar to the number of views for the 'Hello Qiskit' section. During this time, 71% of views for the chapter come from within the textbook. These statistics suggest that it serves an integrated role as part of the introductory materials within the textbook. This motivated the more prominent usage of the visualization in the 'Visualizing Entanglement' section of the new interactive Qiskit introduction course.

These results demonstrate an additional benefit to the design of quantum games and interactive tools. By finding visual and interactive representations of the mathematics behind quantum computing, we make it easier to describe. This allows us to create more engaging and understandable explanations in text. It is therefore worth considering making fully blog based companions to any educational quantum game or interactive tool.

It is worth noting briefly that a similar insight was gained from the game *Battleships with partial NOT gates*. This was one of the first games to use a quantum computer, and was designed primarily so that the source code could be explained in a blog post[133]. There is therefore little evidence of it being played, but the blog post remains the most viewed on the Qiskit blog by some margin.

## 4.2   Particle in a Box / Psi and Delta

*Particle in a Box* and *Psi and Delta* are digital games designed to support high school and undergraduate students in learning introductory quantum physics. The games were designed and developed as part of an interdisciplinary team of electrical engineers, media scholars, and HCI researchers comprised of undergraduate and graduate students as well as faculty at Georgia Tech. *Particle in a Box* aims to support learning quantum physics by allowing students to virtually explore, compare, and experiment with worlds that follow the laws of classical and quantum physics[61–63,134]. Building upon *Particle in a Box*, *Psi and Delta* aims to teach quantum physics through collaborative inquiry by requiring two players to cooperate, develop, and employ concepts of quantum mechanics together in order to resolve challenges involving quantum physics[135]. Both games can be downloaded and played at *https://learnqm.gatech.edu*.

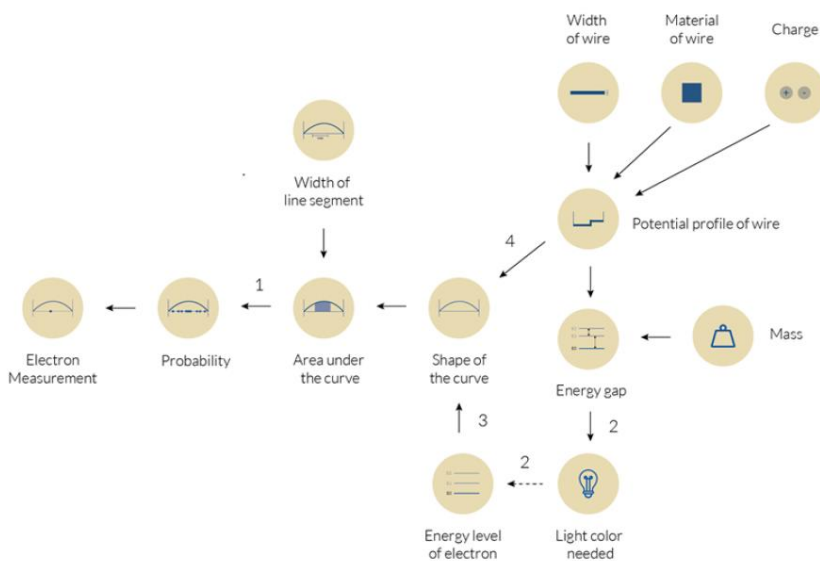

Figure 5. The Concept Map of the square well-problem.

Both games integrate quantum physics into the game mechanics using a concept map of the infinite square well. For example, to reify the concept of light color affecting energy levels, *Particle in a Box* requires students to find and collect light bulbs of the appropriate color and bring them to a lamp to shine light onto an electron. Similarly, to promote engagement with the relationship between "Area under the Curve" (the curve referring to the probability distribution of the electron) and the probability of the electron, *Psi and Delta* uses platforms with different widths and measurement counters that count how many times an electron is measured under each platform to express probability. The concept map of the square well-problem used by both games is provided in Figure 5.

4.2.1 Particle in a Box

*Particle in a Box* is a 2-D single-player platformer game built in Unity where the player plays the game through a virtual avatar who travels through a two-dimensional rendition of the classical and quantum worlds. The classical world, set in a lab-like environment, (Figure 6) involves challenges based on Newtonian physics such as jumping over a ball and increasing its total energy. After playing the classical world, players shrink down to the quantum world (Figure 7) which is based on the infinite square well problem. It involves an electron that is trapped in a 1-D quantum wire and exhibits quantum behavior such as superposition and energy quantization. The challenges here are based on understanding introductory quantum physics. Experiencing both worlds affords students the opportunity to compare their similarities and differences—an approach that has shown to be effective in learning quantum physics.

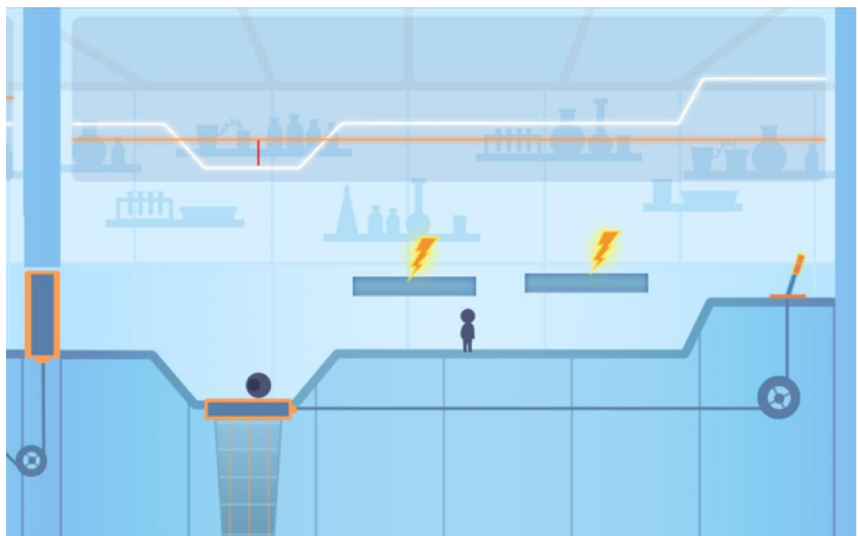

Figure 6. The classical world in Particle in a Box. The player must reach the energy bolts, carry them, and place them in the path of the moving ball.

Both worlds have similar objectives: to raise a particle's energy. In the classical world (Figure 6), the objective is to raise the energy of a rolling ball, so that it pushes a lever located higher up, which opens the door to the next level. In the quantum world (Figure 7), the objective is to increase an electron's energy up 3 levels by gathering light bulbs of the correct color and shining light on the electron's quantum wire. Having similar objectives enables direct comparison between concepts such as position and energy, which manifest differently between classical and quantum physics.

Further, in both worlds, if the player hits the particle, they faint and must restart. In the classical world this means coming in contact with the ball. In the quantum world this means coming into contact with the measurement of an electron particle. The electron by default is in a state of superposition. However, it is also measured regularly and 'appears' each time as a particle. The position where this electron particle will appear is random but follows a probability distribution (curved white line in Figure 6). Through repeated measurements, the game simulates an 'experience' of probability as the player learns to identify the locations where the electron is likely to be measured and therefore where they are more likely to be hit. Since players faint if they are hit by the particle in either world, they must learn to carefully analyze the behavior of those particles in both worlds, thereby promoting further comparison.

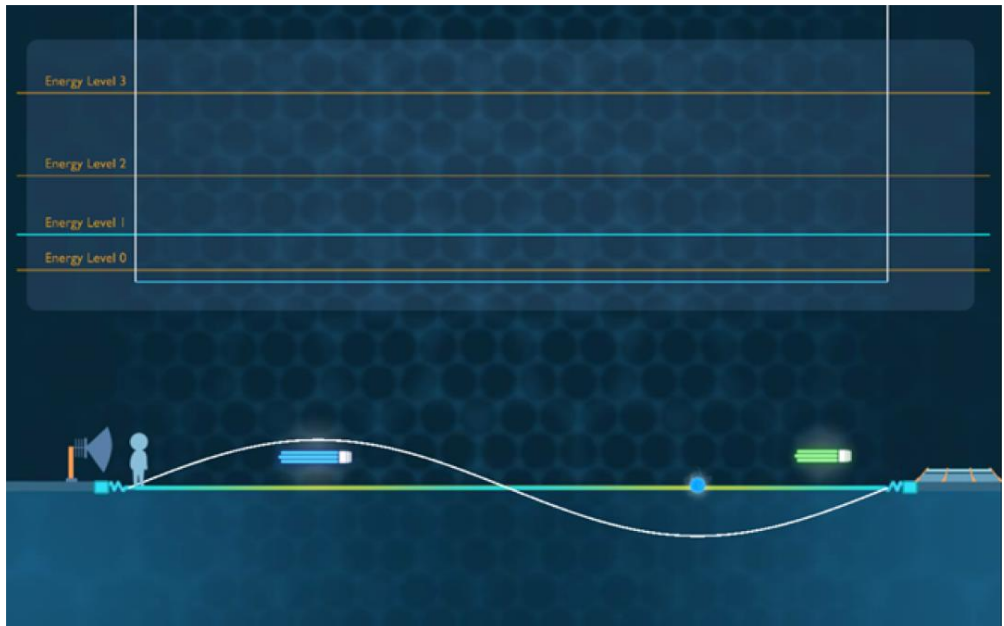

Figure 7. The quantum world in *Particle in a Box*. The player must carry the colored bulbs to the lamp (left) that will shine light and increase the electron's energy level (top). The blue dot is the electron and the bright line it is on is the wire. The curved white line illustrates the wavefunction of the electron.

*Particle in a Box* was evaluated with undergraduate electrical engineering students at Georgia Tech and showed an improvement in students' conceptual understanding of probability[61]. Students further reported an increase in comfort level with the key concepts taught in the game. The game was awarded the Student's Choice Award at the Serious Games Showcase and Challenge in 2015.

4.2.2 Psi and Delta

*Psi and Delta* builds on our findings *Particle in a Box* by students adopt the role of two robots, with the aim of defeating opposing robots in a world governed by the laws of quantum mechanics. Students accomplish this task by using concepts of quantum mechanics to lure and "shock" the opposing bot. If a student's bot touches the opposing bot or gets "shocked", it loses part of its health. If any player's bot loses all their health, the level restarts.

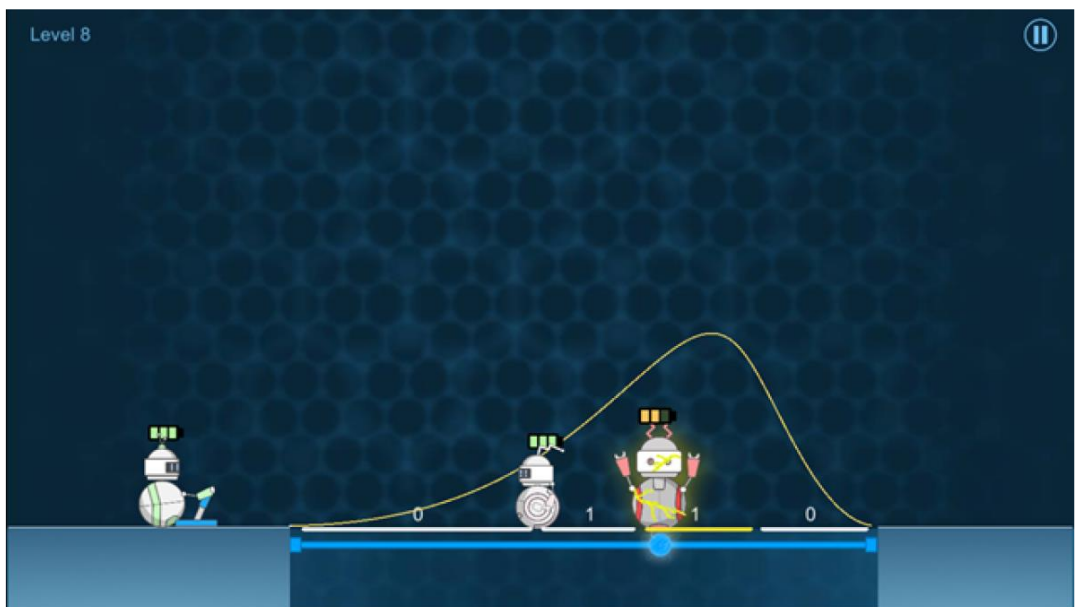

Figure 8. Taking a measurement in Psi and Delta. One player (left) takes the measurement by pulling a lever, while the other (middle) lures the enemy bot to a higher probability area. Here, the enemy bot is getting shocked by the measured electron as it is standing on a platform under which the electron was measured.

The game is divided into two parts. In part one, students develop models of the concepts of superposition and probability (Figure 8). As discussed earlier, when an electron is confined in a small area, it will enter a superposition state, i.e., it will exist in multiple positions simultaneously. To break this superposition, one needs to take a "measurement." In the game, the electron is confined to a small blue quantum wire. In contrast to the automatic measurements, in Particle in a Box, it is the players who take measurements by pulling a lever. Each time a player pulls the lever, a measurement is taken which collapses the electron from its superpositions state to an unpredictable position on the wire for a brief moment. Any robot (player or enemy) standing on a platform directly above the collapsed electron will get "shocked" and lose some health (see Figure 8). The position where the electron collapses is probabilistic, i.e., some positions are more likely than others. The relative probability of these positions is illustrated by the electron's probability distribution, the orange curve in Figure 8. The longer the platform and the higher the curve above it, the more likely the electron will be measured under it. After each measurement, the electron returns to superposition.

In part two, students develop models of energy levels (Figure 9). Here, the opposing bot has a shield which protects it from getting shocked. To break the shield, the electron needs more energy. Electrons can only have a discrete amount of energy such as 1 eV (electron-volt) or 3 eV in the case shown in Figure 9, but nothing in between. Energy can be supplied to an electron in the form of light, which consists of discrete energy packets (photons) whose energy depends on their color. To excite an electron from a lower energy (say 1 eV) to a higher energy (3 eV), one must shine photons with the exact energy as the gap (2 eV). In the game, students can shine light using a lamp and also change its color using a spectrum.

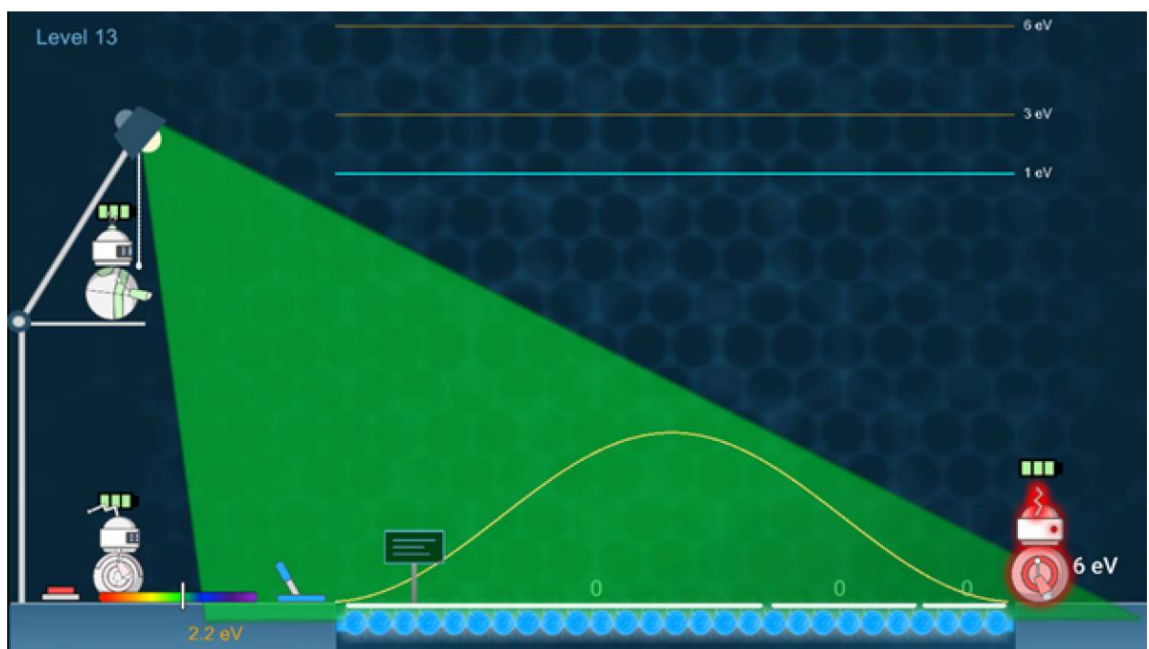

Figure 9. Shining light to increase the electron's energy level in Psi and Delta. The horizontal lines on top indicate the electron's energy levels. The spectrum (bottom- left) lets players change the color of light shined.

This collaborative inquiry approach encourages students to share their ideas and work together to understand basic quantum physics concepts[135]. Qualitative evaluations of *Psi and Delta* with undergraduate Physics and Astronomy students showed that it was effective at engaging them with the basic concepts of quantum physics and had potential to be integrated more formally into traditional lectures[18]. The game was also presented at the Smithsonian Creativity and Innovation Festival held at the National Museum of American History in Washington DC.

## 4.3 QPlayLearn

*QPLayLearn*[136] is an online platform conceived to explicitly address, within a global strategic vision, the challenges and opportunities of widespread quantum literacy. *QPlayLearn* was born from a group of scientists passionate about quantum science and firmly believing that everyone can learn about quantum physics and its applications. *QPlayLearn*'s mission is to provide multi-level education on quantum science and technologies to anyone, regardless of age and background. To this aim, interactive tools enhance the learning process effectiveness, fun, and accessibility, while remaining grounded with scientific rigor. As a strategy for cultural change in a wide range of social strata, *QPlayLearn* offers diversified content for different target groups, from primary school all the way to university physics students, to quantum science enthusiasts. It is also addressed to companies interested in the emergent quantum industry, journalists, and policy makers needing to grasp what quantum technologies are about.

Inspired by the theory of multiple intelligences[137], *QPlayLearn*'s holistic perspective stems from the recognition that different types of intelligence dominate the learning process of each person. Therefore, the platform is conceived to accompany users via different paths aimed at: 1) building intuition and engagement through games and videos; 2) understanding physical concepts through accessible and scientifically accurate descriptions, graphics, animations and experiments; 3) acquiring formal understanding through mathematics. Such a three-step method is used in Quest, the *QPlayLearn* quantum dictionary, where each entry is composed of various sections exploiting different kinds of learning-approaches and resources. For each concept in the dictionary, each user can begin from the approach that feels easier to them, and then possibly explore the others. Eventually, it is the combination of the different pathways that shifts and expands the understanding of quantum physics and technologies.

In this context, games are one of the tools, embodied in the section Play, to stimulate the interactive participation of the users to grasp the counterintuitive features of quantum physics. Besides games, short animations (Quantum Pills) explain concepts at different depths of understanding, useful for experts and non-experts. In the Discover section, specialists explain concepts with short videos, using metaphors, experiments or deductive examples. The Learn section enters the formal core of quantum theory, devoted to a more expert audience. In the Apply section, concepts can be practiced by running code samples on real-world quantum devices. Finally, the Imagine section offers an environment totally free from restrictions, where learning proceeds via the preferred user's artistic language, centering the journey around creativity itself.

Because of the foundational concept, *QPlayLearn* is aimed at wide and diverse (free) use, to tailor education processes on quantum science on the many aptitudes of different users, operate a diffuse and massive cultural change and build up literacy and awareness. Because of this flexibility, *QPlayLearn* can be incorporated by educators in a widespread context. The effectiveness of the *QPlayLearn* approach needs to be assessed by means of physics education research tools in on-purpose designed experiments. Meanwhile, a number of developments are underway, such as completing the quest environments, adding a multiple language version, and extending the class of beneficiaries to 0-99 years old users, together with pedagogy experts. In fact, the latter represents an authentic priority and an unprecedented challenge[18].

## 4.4 Virtual Lab by Quantum Flytrap

Quantum Flytrap is a company founded in 2019, focused on bringing quantum technologies closer to users by developing user-friendly graphical interfaces (GUI). On top of on-going commercial projects involving developing a no-code Integrated Development Environment (IDE) for quantum hardware, Quantum Flytrap provides educational tools to demonstrate capabilities of interfaces and real-time simulations.

*Virtual Lab* by Quantum Flytrap (2019-)[138] is a real-time online simulation of an optical table, supporting up to three entangled photons. The Lab uses a drag-and-drop interface for positioning optical elements such as photon sources, mirrors, polarizing and non-polarizing beam splitters, and Faraday rotators. The inspiration comes from ease of use and capabilities of construction systems such as LEGO Bricks, the Goldberg machine video game The Incredible Machine[139], and grid-based games such as Chromatron and Minecraft. Most elements have configurable parameters such as phase shift (for reflections, delays and wave plates), polarization rotation (for Faraday rotators and sugar solutions), and absorption rate. The *Lab* supports both destructive measurements and non-demolition POVM measurements.

*Virtual Lab* is a successor to *Quantum Game with Photons* (2016)[140], an open-source puzzle game with 34 levels and a sandbox. Quantum Game with Photons simulates a single photon and its polarization. While most setups could be understood within the framework of classical wave optics the binary character of photon detection shows a few quantum phenomena. For example, the game demonstrates the Elitzur–Vaidman bomb tester - an experiment in which the quantum measurement plays a crucial role.

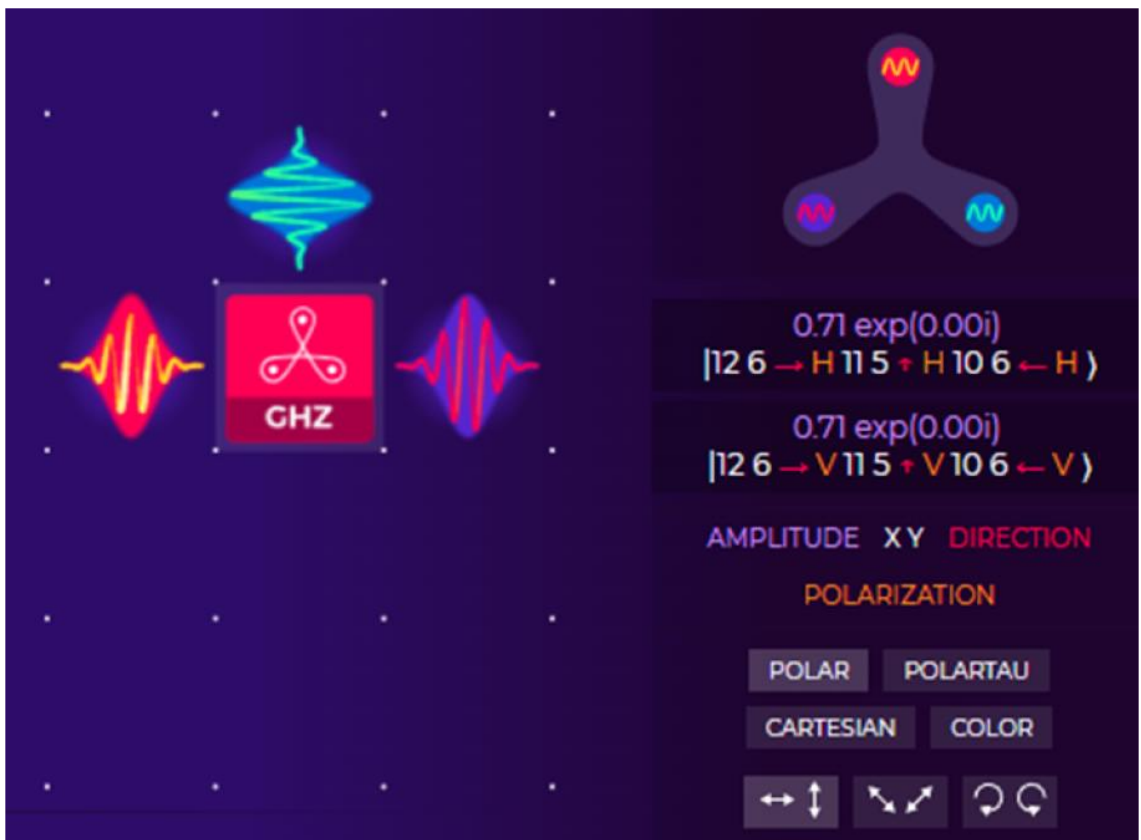

Figure 10. A three-particle Greenberger–Horne–Zeilinger entangled state, visualized simultaneously as (A) waves on the board, (B) entanglement between particles, and (C) the quantum state, in bra-ket form.
22

The original Quantum Game with Photons had over 100k gameplays and was cited in seven peer-reviewed publications by unrelated researchers[18,141–146]. It was selected as the top pick in gamifying quantum theory in The Quantum Times[147]. Based on in-person and email feedback, we learned that it was possible to solve a large fraction of levels without prior exposure to quantum physics, e.g., for high school and undergraduate students. At the same time, a few levels provided a challenge to PhD students and established quantum optics professors.

*Virtual Lab* is focused on reinventing the interface, design, and numerical performance. Our simulation supports multiparticle quantum physics. Unlike most other simulations, it goes beyond qubits. Virtual Lab simulates pure quantum states with photon polarization, four cardinal directions, and a discrete position of a photon on a grid. Consequently, for a typical grid size a single photon requires around 1000-dimensional vector space, so a three-photon simulation requires a billion dimensions. Therefore, Quantum Flytrap needed to develop a custom high-performance sparse array simulation in the low-level programming language Rust - as none of the off-the-shelf solutions were suitable.

*Virtual Lab* offers a set of example experimental setups demonstrating key experiments in quantum measurement, quantum cryptography, quantum communication, and quantum computing. Example setups involve single photon experiments such as nonorthogonal state discrimination, the Elizur-Vaidman bomb tester, the BB84 quantum key distribution protocol, and the concept of non-demolition measurement affecting interference. The two-particle setups include the Ekert quantum key distribution protocol, the Deutsch-Jozsa algorithm, and the CHSH Bell inequality violation. With three particles we can demonstrate quantum teleportation and Greenberger–Horne–Zeilinger correlation.

The measurement results can be subject to logical operations (such as AND or XOR), used to alter other elements (e.g., so to perform quantum teleportation), treated as a goal (for puzzles and course assignments), correlated (for Bell inequality violation), or saved as a CSV table for arbitrary processing.

Moreover, a laser mode allows us to explore classical wave phenomena, such as the three-polarizer paradox (generalizing to the quantum Zeno effect), interferometers (Mach-Zehnder, Michelson-Morley, Sagnac), optical activity, and magneto optic effect (with applications such as an optical diode).

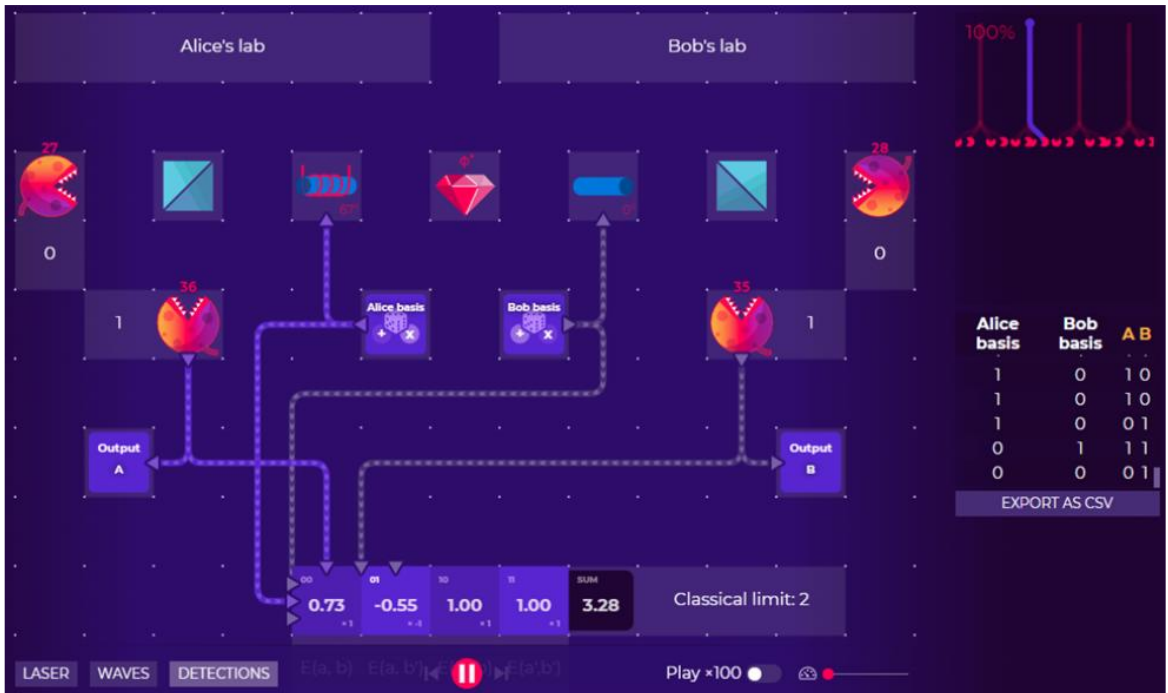

Figure 11. An experimental setup for the CHSH Bell Inequality violation. Note that the experiment depends on two random variables (Alice's and Bob's basis), provides a table of all measurements (downloadable as a CSV file) and presents all correlation in real time.

*Virtual Lab* provides multiple ways to investigate an experiment, its outcomes, and the current quantum state. The multiverse tree feature allows users to explore the whole experiment, including all possible measurement scenarios. In the Copenhagen interpretation, all branches are related to probabilistic outcomes. Within the Everettian many-world interpretation, each branch is a coexisting part of the quantum state, representing worlds that are unlikely to interfere with each other.

In each time step it is possible to explore the exact state (ket) with a preferred choice of the basis and coordinates of complex numbers. Furthermore, we visualize entanglement of every particle with the system.

*Virtual Lab* has been used for Quantum Information courses at Stanford University and the University of Oxford, as well as by Qubit by Qubit's Introduction to Quantum Computing online course. *Virtual Lab* features over 450 custom, user-created setups. It has around 70 users on a regular working day, and peaks at over 700 during events.

Quantum Flytrap released *Quantum Tensors*[148] - an open-source JavaScript library for in-browser quantum information processing. It goes beyond qubits and supports physical multiparticle physical systems (polarization of light and spin), as well as qudits. It allows the creation of states, change bases, perform unitary operations, measurements (destructive and non-demolition POVMs), and entanglement entropy. It has sparse linear algebra operations (including with tensor products), as well as support for quantum circuits.

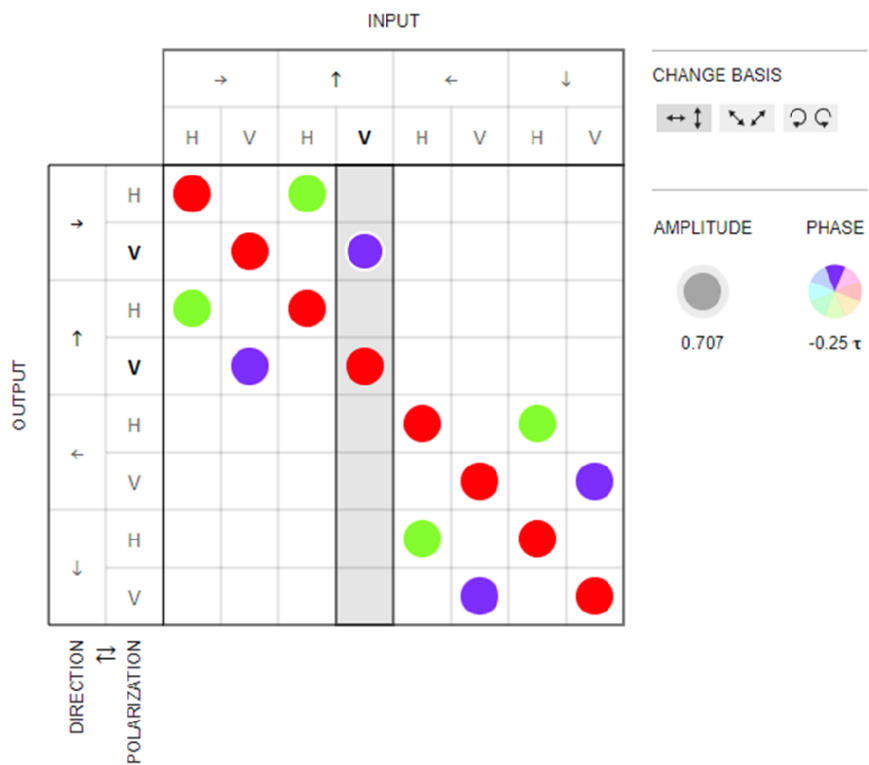

Figure 12. A matrix visualization for a beamsplitter with BraKetVue. The matrix values are presented with a circle size (radius for the amplitude, area for the probability) and color (hue for the phase). Index labels incorporate the tensor structure.

*Virtual Lab* allows exploration of quantum states at each single time step (see Figure 11), as well as matrices for unitary operation and projections (see Figure 12). Quantum Flytrap open-sourced BraKetVue (or ⟨φ|ψ⟩.vue)[149], a visualization of quantum states and operations in the popular web-development framework Vue.js. It allows to create interactive quantum explorable explanations (e.g. in a blog post with an introduction to quantum gates[113]), and can be used in other web-based environments, such as presentations with Reveal.js (created in RISE in Jupyter Notebook or with R Markdown).

Matrix visualizations allow us to show magnitude and phase of transition in a visual way, dynamically change basis, and explore the tensor structure (e.g., change the order of particles).

## 4.5 Quantum Odyssey

*Quantum Odyssey* is a privately-funded, large software project, still in development by Quarks Interactive[150], but the alpha version is already available for the public. Currently the software comes with over 20 hours of fully narrated, self-paced education ranging from quantum optics phenomena to coding quantum algorithms. The software also contains a vast encyclopedia that verbosely explains key quantum mechanics concepts. People working on the software have various backgrounds (researchers in education, quantum physicists, artists and game developers).

Behind *Quantum Odyssey* is the concept of quantum literacy, as coined by Nita et al.[19]. The aim of *Quantum Odyssey* is to make quantum physics and computing knowledge accessible to all audiences, no matter their background, by making learning about and working with quantum algorithms exciting and fun. *Quantum Odyssey* is a fully visual and software-assisted method to learn how to create new and optimize quantum algorithms for quantum computers. Equivalent to everything that can be achieved using the gate model framework, users can design quantum algorithms for any purpose (ranging from designing physical interactions in optics to more advanced quantum computing algorithms) without requiring any background knowledge in STEM. *Quantum Odyssey* shows in real time the dynamics of the quantum state vector, including its evolution as the user adds quantum gates to build quantum algorithms (Figure 13).

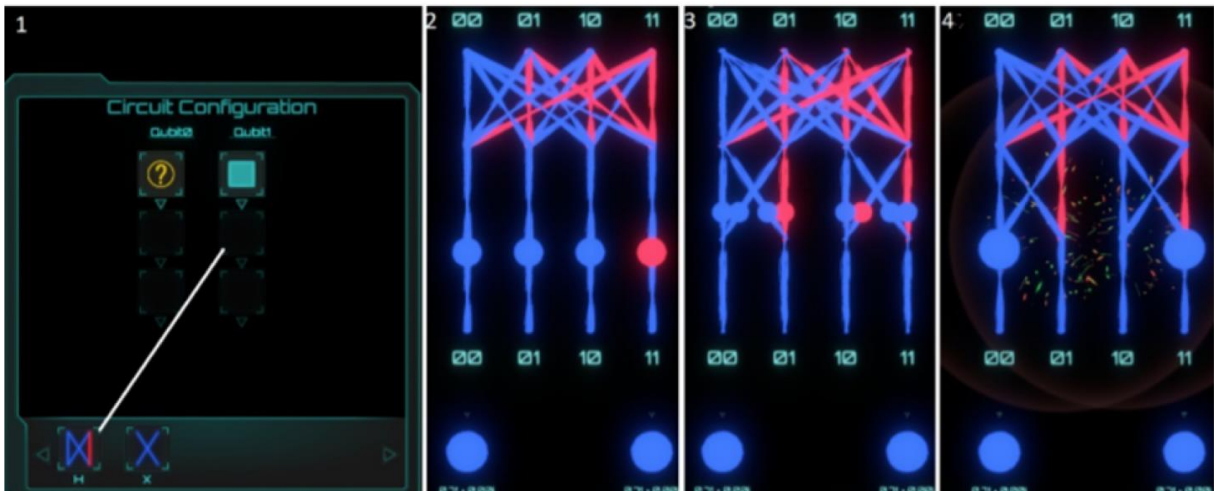

Figure 13. Images 2, 3 and 4 are snapshots of the evolution of a quantum state vector in visual form for 2 particles that start in the ground state 00 and reach the Bell state 00 and 11. The balls at the bottom show the desired output solution and the user is required to place a Hadamard gate in order to achieve it as shown in 1.

The software was tested in schools and found to be suitable for teaching quantum physics and computing from the age of 12 and above. The results of the research study[117] show that the *Quantum Odyssey* visual methods are efficient in portraying counter-intuitive quantum computational logic in a visual and interactive form. This enabled untrained participants to quickly grasp difficult concepts in an intuitive way and solve problems that are traditionally given today in master's level courses in a mathematical form. The results also show an increased interest in quantum physics after play, a higher openness and curiosity to learn the mathematics behind computing on quantum systems. Participants developed a visual, rather than mathematical, intuition that enabled them to understand and correctly answer entry level technical quantum information questions. Another interesting find is that the design of the visuals of the software makes it equally appealing to both female and male participants. That is, both genders were found to be equally able to solve complex mathematical problems if they were given in a visual puzzle form within *Quantum Odyssey*.

In future releases, the following new features will be introduced:
- *Quantum Odyssey* will provide the ability to import/ export visual puzzles to IBM Qiskit and run algorithms right on an IBM QCPU. Everything designed with *Quantum Odyssey* is fully compatible with Qiskit.
- Multiplayer, collaborative efforts in optimizing and discovering new quantum algorithms.
- Community puzzle publishing and ranking system for puzzle creators and puzzle solvers.

## 4.6   ScienceAtHome

*ScienceAtHome*[151] was founded in 2011 at Aarhus University with the goal of developing gamified tools to enable citizen scientists to participate in cutting-edge quantum research. Today, *ScienceAtHome* boasts a suite of games and tools for education and citizen science in a variety of academic fields.

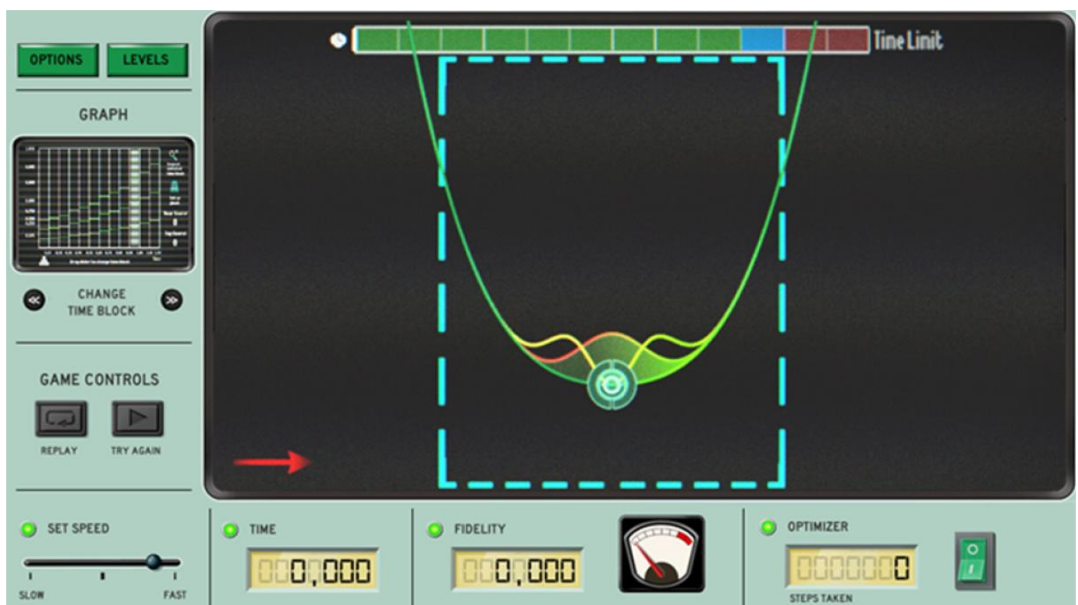

Figure 14. The Quantum Moves 2 interface, showing the *Shake Up* level. The goal is to move the potential (green line) left and right within the cyan dashed box to excite the quantum fluid (green shaded curve with red-yellow outline) from its initial, ground state (as shown) to the final, first excited state (yellow outline). This must be done within a certain amount of time, and the goal is to achieve the highest overlap (fidelity) between the final state and the yellow-outlined state. For more information, see[116].

*ScienceAtHome*'s flagship game is *Quantum Moves 2*[152] (Figure 14), a game where citizen scientists can play through a variety of scenarios, including three scientific problems relevant to cutting-edge research in quantum technologies. To-date, at least 250,000 players have played

*Quantum Moves 2.* Research has shown that player-generated solutions could have some advantages over random seeding when applied to optimization of the scientific problems, even when the solution landscapes of these problems is complex[116]. In particular, players explore the solution landscape more efficiently and can capture solution strategies that algorithms may miss. This is not to say that players are better than algorithms, but rather that the insights of citizen scientists can be useful when deciding how to solve a given quantum problem.

In 2017, researchers at *ScienceAtHome* opened up their quantum laboratory to the general public in the *Alice Challenge*[153]. The goal of the *Alice Challenge* was for users to optimize the number of rubidium-87 atoms in a Bose-Einstein condensate, or BEC[154,155], a cloud of extremely cold atoms that can then be used for a variety of quantum technology experiments. We found that citizens operating the experiment remotely[156] were able to create larger BECs than the local experimental experts[157], a difference that we attribute largely to interfacing and gameplay (Figure 15). That is, we built a simple and intuitive interface that allowed citizens to collaborate (in that any player could copy a previous solution and modify it) to solve the problem-at-hand; such intuitive interfacing is a key for success when considering citizen science projects, quantum or otherwise.

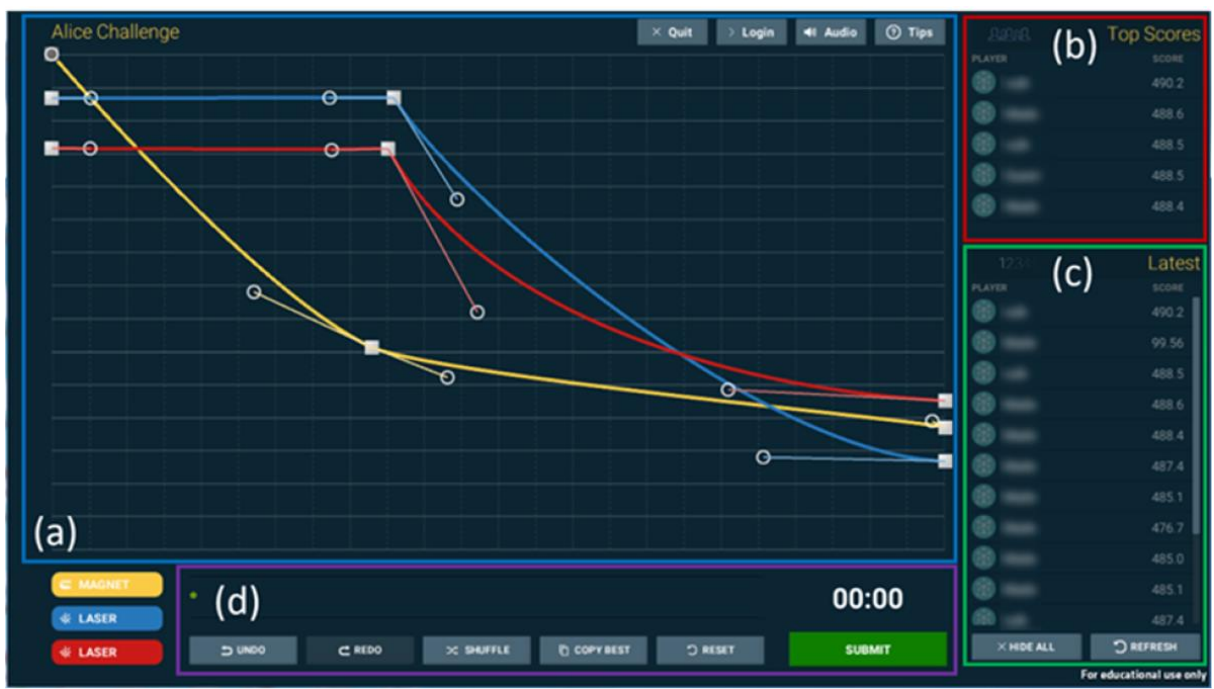

Figure 15. The interface for the Alice challenge, showing (a) the spline curves that the users could manipulate to control the time-dependent laser powers and magnetic field coil currents that controlled the final BEC atom number, (b) a top score list, (c) a list of the latest submissions and their scores and (d) a control toolbar, including the estimated wait time. Players could take any other player's solution (as shown in (b) and (c)) and modify it. Figure reproduced from[156].

*ScienceAtHome* has also developed a number of educational tools, including the *Quantum Composer*[85], which is built on the *QEngine* C++ library[158]; both tools are freely available for download and use. The *QEngine* is a research tool for the simulation and control of one-dimensional quantum mechanics[159], and the *Quantum Composer* wraps the functionality of the *QEngine* into a flow-based tool for quantum visualization. Thus, the *Quantum Composer* (Figure 16) is aimed at introducing students to quantum mechanics and quantum research via a modular interface that can be used to simulate a wide range of problems in introductory and advanced quantum mechanics, including those that are not analytically tractable[86,87].

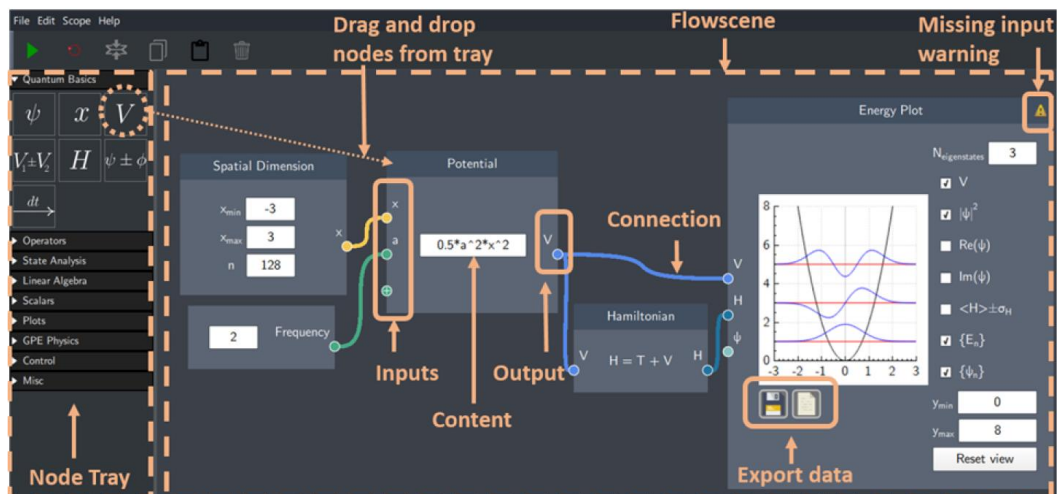

Figure 16. The Quantum Composer interface. Users can drag-and-drop *nodes* (representing space, the potential, a plot, etc.) from the left-hand tray into the simulation space (called a *flowscene*). Connections are then made between the nodes to direct information from one node to another. This figure shows how one can set up a simulation that shows the first three eigenenergies and eigenstates of the quantum harmonic oscillator. Figure reproduced from[86].

*ScienceAtHome* researchers and educators have used these tools in a number of educational and outreach settings, including the aforementioned week-long *Alice Challenge* in 2017 where 600 citizen scientists around the world took part. *Quantum Moves 2* featured heavily in the *2018 ReGAME cup*[160] where more than 2000 students from 200 high school classes around Denmark played a set of citizen science games as they worked through educational content that scaffolded the games, and the planned *Project FiF*[161] (Forskningsspil i Folkeskolen, or Research games in high schools) will expand this to 150 schools around Denmark. *Quantum Composer* has also been widely used as a part of *ScienceAtHome* outreach efforts in Denmark and around Europe, including a three-year-long outreach training effort coordinated as a part of the quantum sensing and control (*QuSCo*[162]) Marie Skłodowska-Curie interactive training network[163]. The outreach training program was aimed at teaching PhD students to communicate their research via a variety of methods, including videos, blog posts, and the development of inquiry-based learning sessions.

## 4.7 The Virtual Quantum Optics Laboratory (VQOL)

The *Virtual Quantum Optics Laboratory* (VQOL) is a graphical web-based software tool for designing and simulating quantum optics experiments[164,165] developed in the Applied Research Laboratories at the University of Texas at Austin. The VQOL was developed initially as an educational tool for introducing high school and early college students to quantum information science, but it has also proven to be useful as a research tool for analyzing real-world experiments[166]. The graphical interface presents a top-down view of a notional optics table with a menu of components that can be placed and configured as desired. To aid visualization, we use six base colors to represent polarization and interpolate these across the Poincaré sphere (See Figure 17). A variety of components are available, including passive linear optics (waveplates, polarizers, beam splitters, etc.), light sources (LEDs, lasers, and Gaussian entangled states), and measuring devices (power meters and detectors).

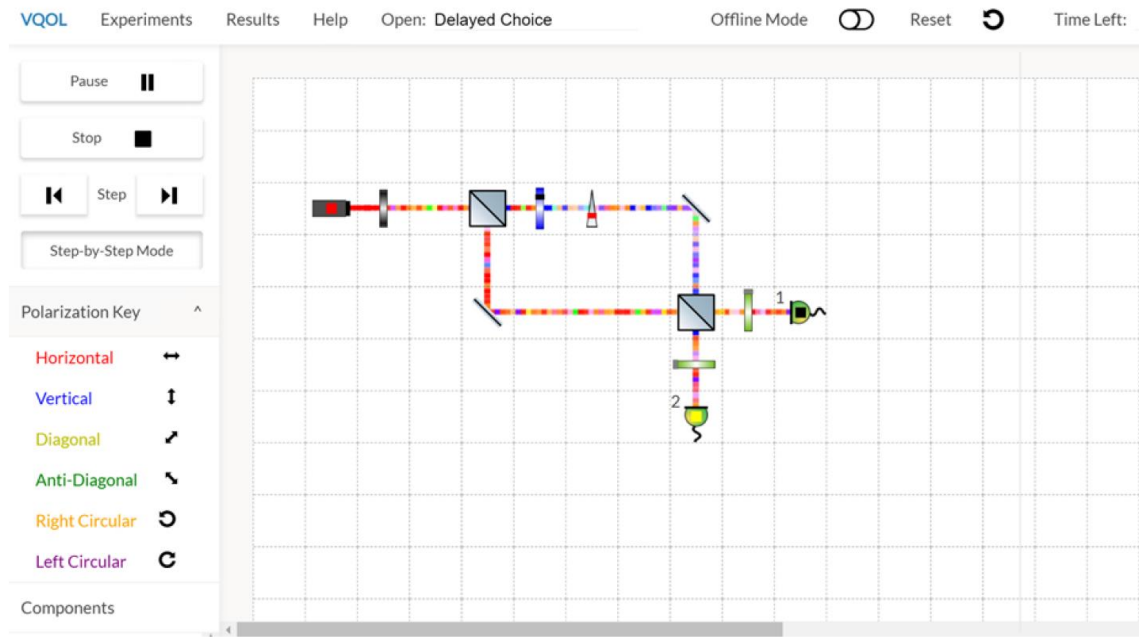

Figure 17. Screen shot of the VQOL interface showing a delayed-choice experiment. A laser (left) produces horizontally polarized light (red) that is attenuated by a neutral density filter before entering a Mach-Zehnder interferometer. The random polarizations (multiple colors) are due to vacuum noise. The upper arm has a half-wave plate (blue disk) that can provide which-way information by changing the polarization to vertical (blue). The two detectors have polarizers (green disks) to detect horizontal light.

Unlike traditional quantum simulators, VQOL works within an entirely classical framework, with vacuum modes represented explicitly as complex Gaussian random variables. Detectors operate deterministically and "click" when the local amplitude falls above a prescribed threshold, which also determines the dark count rate. Thus, the only fundamental source of randomness lies in the vacuum modes, which are independently generated at each time step. By default, time steps are taken to be 1 µs long and correspond to both the coherence time of the sources and the dead time of the detectors. Quantum phenomena arise from the use of non-Gaussian measurements (i.e., detectors) and post-selection. This unique combination of features allows for a subtle exploration of the quantum-classical boundary.

In an educational setting, VQOL may be used to introduce optics concepts (either classical or quantum) or to explore quantum phenomena and applications. Over the past two years, we have used VQOL as part of a full-year high school course on quantum computing[167] as well as part of a two-semester introduction to quantum computing for university freshmen and sophomores. A typical introduction to qubits begins with the concept of polarization, which can easily be understood in terms of transverse waves on a string. We then introduce projective measurements through an initial investigation of Malus's law using a laser, polarizer, and power meter. We transition to the Born rule by introducing a neutral density filter, which attenuates the laser light to the single-photon regime, and replacing the power meter with a detector for photon counting. Students perform guided experiments and explore the phenomena before drawing inferences to uncover the physical laws behind them. Upon this foundation of physical intuition, a more concise abstract mathematical description may then be introduced. One important aspect of these experiments is that students are challenged to consider appropriate data analysis methods for conceptually connecting "counts" with probabilities. Since, from an experimental point of view, there is not a well-defined number of "trials," this exercise raises interesting interpretive questions. Students must consider how to normalize their data and account for dark counts, just as they would need to do in a physical optics lab.

Beyond introducing elementary concepts, we have also used VQOL for deeper explorations into quantum physics. Over the past two years, we have offered a virtual experimental quantum optics program to our university students. The program aims to connect theory to experiment by exploring quantum information science in an experimental optics setting. Students work in groups of 2-4 individuals and are tasked with exploring about 10 different experiments over a 9-week period. Students select which experiments to investigate and are given only general guidance on the goals and phenomena to be explored. At the end of each week, the groups get together to present and discuss their findings. Possible experiments include quantum state tomography, an optical implementation of Deutsch's algorithm, delayed choice and quantum eraser experiments, quantum teleportation, entanglement swapping, and violations of Bell's inequality[164]. We have found through post-program surveys that students generally respond well to the open-ended nature of the experiments and feel that it gives them a more authentic research experience than traditional, more structured undergraduate laboratories.

## 4.8   Activities in Creating Quantum Games

The first recorded hackathon was organized by developers of the OpenBSD[168] software in June 1999, and the term was coined by its organizers. It is commonly used to refer "*a usually competitive event in which people work in groups on software or hardware projects, with the goal of creating a functioning product by the end of the event.*"[169] The first (or the $0^{th}$ depending on perspective) *game jam* was organized by *The Indie Game Jam* community in March 2002[170]. It usually refers to a more relaxed and non-competitive event compared to a hackathon, like a jam session of a musical band. Ever since their emergence, both terms were adopted by software communities and beyond, and the number of hackathons and jams grew to hundreds per year[171]. Therefore, their adoption by the community of quantum game developers was not a surprise.

The first quantum game jam was organized in Finland in 2014 as a part of the *Finnish Game Jam*[172] and the first quantum hackathon (not focused on games) was organized in April 2018 by Rigetti[173], where two quantum games were also developed. Following these, several quantum game jams[174] and hackathons focused on game development were organized.

In what follows, we provide a recent example of a quantum games hackathon organized in September 2021. This is followed by the description of a quantum computer games project-based course at the University of Texas at Austin, organized by the developers of the previously introduced *Virtual Quantum Optics Laboratory* tool. Both examples are described by their organizers in detail, to be used as templates by any interested readers in our audience.

### 4.8.1 Quantum Games Hackathon by Quantum AI Foundation

The *Quantum Games Hackathon*[175] was an online event organized in 2021 by the Quantum AI Foundation (QAIF)[176] in collaboration with members of QPoland[177], a QCousins branch of QWorld[178]. The event was related to the Warsaw Quantum Computing Group meetups[179]. The main motivation of this hackathon was to popularize and foster innovation in quantum computing as well as to provide for the participants an opportunity to develop quantum programming skills. The hackathon started with two days of introductory talks and workshops (25-26th September 2021) followed by a week for project development. The submission deadline, as well as the announcement of the winners, was the third of October 2021. Additionally, QAIF and QPoland organized a separate workshop prior to the hackathon based on QWorld's Bronze[180] material in order to give participants more time to learn necessary skills.

The invitation to the event was shared on social media platforms: Facebook, LinkedIn groups, and Twitter. There were 128 registrants from 41 countries. The participants were working in teams composed of up to five people, formed during a match-making event. Participants were advised to have team members with expertise in different domains, e.g., quantum programming, user experience design, business development, computer games development, etc. Communication within teams, with mentors and with the organizers was held using the hackathon Discord server. Besides the quantum computing workshop, participants had an opportunity to listen to lectures given by experienced researchers, developers and entrepreneurs working in the quantum computing domain. The talks, given over two days at the beginning of the hackathon, included introductions to software and programming platforms (*Qiskit* and the *JavaScript* environment), existing games (e.g., *Quantum Odyssey*), and interface design and accessibility in games. All lectures were given using Zoom, most of them were recorded and are available on QAIF's YouTube channel[181].

At the end of the event, seven solutions were submitted. According to the hackathon rules, the participants were requested to submit a link to a public GitHub repository containing the quantum game code and a short video presentation (up to five minutes). All the submissions were assessed by a jury composed of six members according to the following criteria:

- **Correctness:** Are the assumptions and usage of quantum mechanics and quantum computing appropriate?
- **Playability:** Is the game entertaining? Does it have educational potential? Is the user interface inviting?
- **Originality:** How creative and innovative is the idea?
- **Quality and completeness:** Is the game polished and ready to play? Are the mechanics well introduced?

Three best projects were awarded with prizes, there was also an additional prize funded by *Quantum Flytrap* for the best game developed using *JavaScript*. All the other prizes were awarded by the other Strategic Partners of the hackathon: Snarto, Cogit, Quantumz.io and KP Lab. The winning solutions were:

- **Dave the Hackerman by Dodo Boys**, a game on diagrammatic ZX-calculus[182].
- **QuanTetrum**, a Tetris-inspired game that uses 3 important concepts of quantum physics: superposition, entanglement, and measurement.
- **QG Crusher**, a Candy Crusher-like game with quantum gates.
- **Quaze**, a quantum maze in which each maze tile is a gate changing the state of a quantum particle.

4.8.2 Quantum Computer Games project-based course at the University of Texas at Austin

For the mid-semester project in the Quantum Computing stream of UT Austin's Freshman Research Initiative (FRI) seven groups of 1-3 students were tasked with programming their own games to run on quantum computers. In a short lesson introducing the concept of quantum games, we briefly covered motivations for designing games including showing off or testing state-of-the-art hardware, engaging participants for the purposes of education and entertainment, and, importantly, as a creative exploration of new technologies. For the purposes of the project, we defined a game as something with a goal, rules, and feedback. The player uses the rules to attempt to accomplish the goal; feedback is received and, depending on the particular game, may be used to modify the player's strategy. To avoid overly contaminating the students' creativity, we added only that each game must contain four essential quantum effects: superposition, interference, entanglement, and measurement. All these quantum features must be present for the player to manipulate and/or experience.

When designing quantum computer games, students must consider each of the four required quantum effects from a creative perspective rather than simply in an academic context. Using Qiskit, our class builds many of the popular quantum algorithms including Deutsch-Jozsa, Grover search, phase estimation, and factoring; though our teaching style tends to be exploratory, we eventually lay out the algorithms in easy-to-follow steps. The students must understand the steps to apply them in a computer program and analyze the results. In this game-making project, the students do not have concrete steps to follow. Instead, they must either fit quantum effects into a pre-existing game or build original rules around the desired quantum effect. This takes comprehension of quantum concepts as a prerequisite and pushes the student beyond simple understanding into experimentation and imagination.

Each group submitted their programmed game along with a written report highlighting the quantum aspects of their game, potential strategies, and any other interesting features. They also each gave a brief (5-7 minutes) presentation and gameplay demonstration. The game titles and short descriptions are included below.

- **Qhess** - Chess in which each piece can move in superposition. At each turn, all the potential states of a piece must be considered when moving that piece. Measurement occurs when pieces interact by occupying the same square.
- **Quantum Mafia** - A quantum version of the party game, Mafia. Each villager chooses a decimal number as their phase. The mafia member uses quantum phase estimation to attempt to kill the villager. The mafia RSA encrypts their identity, and the detective uses Shor's algorithm to attempt to discover them. The doctor uses the Deutsch-Jozsa algorithm to attempt to save targeted villagers.
- **Quantum Boss Battle** - You must battle a boss monster whose health is represented by a random entangled circuit. Each turn, you attack the monster by applying quantum gates. Your goal is to reduce the state to all 0s. A measurement of 1 or the time running out will result in Game Over.
- **Battle of Millenniums** - The Deutsch-Jozsa algorithm run on a function of your username determines whether you fight for the Angels or the Devils. You have three attacks of varying strength in superposition with a healing action. You also may use a magic attack if you have the magic points available. Your goal is to reduce the enemy to zero health, but damaging the enemy also gives them magic points. Bonuses are awarded according to the outcomes of Grover's algorithm and arithmetic operations in quantum Fourier space.
- **Quantum Uno** - A quantum version of the popular card game, Uno. Each card is implemented as an oracle for Grover's algorithm. Superposition cards take the place of Wilds, Entanglement cards act as a sort of delayed Skip, and a whole new Add Phase card rotates the state of the deck.
- **Quantum FPS** - This is a first-person shooter style game where health is given by the computational basis representation of a 3-qubit state. Hadamard and CNOT gates act as ranged attacks and X gate is a close-range attack. You can change your own state by using H or CNOT on yourself. You can measure yourself or your enemy, but you die if your state is ever measured as |000⟩.
- **Quantum Invasion** - Player 0 and Player 1 compete for territory across a neutral zone which starts in superposition. Each player has a number of actions to spend on different attacks such as rotating a single qubit, rotating multiple qubits, and entangling target qubits in superposition. At the end of the game the state of the board is measured, and the player with more qubits in their state wins.

# 5 The pilot project QUTE4E

Storytelling in quantum outreach is crucial for a number of reasons, including the unintuitive nature of quantum mechanics, but it is also extremely timely if one considers the potential effect that quantum technologies may have on future society. Counterintuitive examples of quantum phenomena (e.g. tunneling and entanglement) provide an engaging means to practice the process of scientific thinking with the general public using creative, experimental and formal literacies. Finally, as a currently developing field, quantum technologies allow educators to potentially integrate Responsible Research and Innovation (RRI) dimensions into a given outreach activity.

The pilot project Quantum Technologies Education for Everyone (QUTE4E) was conceived to respond to the challenges described above by building tools suited to address and measure what aspects outreach and education on quantum technologies should possess to be efficient and effective.

The key concept of QUTE4E is that engagement is a fundamental requirement of education, no matter the learner's technical skills. The idea is that every citizen should be made aware of their ability to access the essential concepts of quantum science irrespective of age or socioeconomic background. Thus, the goal of the pilot is to produce basic and engaging storytelling of the essential concepts in quantum technology in a captivating and language-accessible manner, designing suitable resources to educate intuition and boost creativity for every target audience: the public, students, educators, school and academic teachers, outreach experts, QT companies or policymakers. Building on this basic conceptual map, the pilot is developing additional ideas and tools with a formal language which is — within learning/teaching/training processes — characterized by an increasing level of complexity and technical content associated with audiences with increasing levels of formal education and corresponding educator training.

Now that we have introduced the pilot project, we describe how these challenges are addressed by the pilot efforts. We can illustrate the importance of storytelling in QT with the example of light.

Light is a transversal concept in physics, and its concepts are taught in subfields of physics ranging from quantum mechanics to cosmology. The story of light and photonics provides a means to engage with a broad audience as it can be tuned from basic concepts, like an explanation of how rainbows are formed and why the sky is blue to specialized technical fields such as quantum optics that directly impact quantum technologies. In addition, since light is familiar to everyone, describing some of its intriguing properties can also encourage curiosity regarding what is behind the concept of light and how to apply it in our daily life, fostering an interest in science and its applications.

Thus, we believe that teaching through storytelling is the right tool to ignite passion and bring excitement and engagement about nature, and, more specifically, the world of quantum technologies. This method allows the audience to take an active role in the search for the explanation for a previously misunderstood observation, exercising scientific thinking and exploring new forms of literacy.

In addition, this approach is particularly applicable to gamified and interactive activities that support the active role of the learner in a natural way by introducing relevant concepts through increasingly difficult challenges and levels, creating natural environments for visualization and player experimentation, and providing feeling of success that can foster increased motivation for further education in quantum physics and technology.

The QUTE4E group is designing a general format for storytelling in quantum science and technology that makes full use of the process of scientific thinking[178]. This format starts from observation to create understanding and is then formalized and finally benchmarked back into fact-checking. In this process, the experimental and formal literacies that are required to perform the different steps are built up using interactive tools like those described in this work. Examples include virtual experimental labs and quantum games, popular science animations and artistic immersive experiences. Once the functional scientific thinking process is scaffolded, one may ask whether engagement is generated through the storytelling of basic quantum science and technologies. In fact, our answer to this question runs along two different dimensions: an educational consideration and a Responsible Research and Innovation (RRI) analysis.

On one hand, we consider that attractive storytelling of both basic quantum science and quantum technology concepts widens the potential audience of learners, bringing about the opportunity to promote problem-solving perspectives via education. Here, we use the functional scientific-thinking process described above both for basic concepts and for more technical ones, where a complex question is fragmented into simpler pieces whose understanding involves basic quantum science concepts. The importance of passing this scientific thinking framework to our audience is paramount.

Finally, we strongly believe that our storytelling should be guided by and articulated within the RRI dimensions, i.e., public engagement, open access, governance, gender equality, ethics, and science education. The QUTE4E group is developing outreach concepts with strong focus and involvement of the stakeholder groups to identify the current needs and to respond with the innovative solutions.

The conceptual map of QUTE4E can be viewed as a HyperSyllabus, composed of six parts times six types of beneficiaries' times six RRI dimensions. The parts are: a dictionary of essential quantum concepts, choices of language and methods, applications of concepts, learning outcomes, assessment methods, resources. The beneficiaries are general public, students 0-18, students BS/MD/Ph.D., their (for each student's grade) educators/teachers/trainers, policymakers, QT re-trainees. Finally, the RRI dimensions are: gender (balanced gender presence and representation in quantum technologies), open access (free resources), public engagement (also via outreach), ethics (what the quantum technologies are for), scientific education (formal and non-formal education to a scientific, synthetic, and creative mind), and relationship with governance (impact of quantum technologies on citizens' everyday life).

The group uses a storytelling approach to foster a better connection to the beneficiaries; as every story creates opportunities for motivation, appreciation, and education while being able to adapt the content and format by being aware of differences in objectives, needs and responsibilities of RRI beneficiaries. Interactive and gamified tools provide an excellent framework for covering the specific needs to help establish target group-related channels for successful communication.

The pilot resources are conceived to be dynamically updated from existing ones. A non-exhaustive list of existing resources developed by the participants to the pilot consortium includes the following: quantum games with a purpose of education (as in *ScienceAtHome*, *QPlayLearn*), also developed in Quantum Games Jams and Hackathons, quantum physics labs (e.g., *Quantum Composer, Virtual Lab by Quantum Flytrap, 3D space, Labster*[183]), quantum concepts labs (*Quantum Composer*), quantum algorithms labs (*IBM Quantum tools/ Qiskit, Quantum Odyssey*), quantum computing resources (*IBM Quantum/ Qiskit*), educational tangible objects for K12 experience activities (*Trump cards at Birmingham*[184,185]), quantum pills and animations (*QPlayLearn, Quantum Tour*[186]), artistic (visual, audible, touchable) immersive experiences (e.g., *Quantum Garden, Quantum Jungle*[187,188]) and online courses for different types of beneficiaries (*QWorld, QPlayLearn, IBM Quantum/Qiskit*). A selection of these has been discussed in Section 4.

Within the bounds of the pilot, a special emphasis is planned on developing methods of extending the 'reach' of a given initiative. Many (already existing) resources lack the proper channels and means to disseminate and be used widely by their potential beneficiaries. As the pilot develops its tools, it can also develop/improve means of reaching target audiences, looking for strategies aimed at making the resources suited to diverse users. Basic outreach activities (such as opening social media accounts or creating online content with no focus on dissemination) are insufficient and fail to reach a wide population. Therefore, the pilot utilizes a wide range of collaborations, together with the informed feedback from a Delphi study, to create, test, and improve a network of dissemination for quantum education-related materials, which will be beneficial to the entire community. Overall, the pilot represents a backbone on which the diverse resources created by the partners can be synergistically made available to boost outreach and education.

In summary, the planned actions include:

- The production of an essential syllabus of quantum physics concepts and their main applications to quantum technologies, identifying for each the type of beneficiary, already-existing educational resources and a list of the missing ones, the learning outcomes, the best-suited type of symbolic/language system, relevant assessment tools, and, coherently with the RRI dimensions, potential applications that the pilot can focus on including public health, the environment, and communication.
- A Delphi study with QTEdu participants and associated parties aimed at identifying stakeholders, opportunities and problems in QT outreach and engagement activities, as well as practical solutions. The study outcomes are expected to represent a guide for future design of outreach and educational resources that can be disseminated to diverse and extended audiences.
- The organization of joint outreach virtual activities, including live panels for the European Researchers' Night, hackathons, quantum game jams, showcases for games and virtual experimental and simulation platforms and a joint program of events for the World Quantum Day.
- Assessment of the coherence of the pilot deliverables with RRI dimensions, which is achieved by defining a checklist for each individual dimension and specific type of resources and activities that define the six-by-six matrix described above, and finally applying the checklist to the pilot deliverables. This checklist will be produced via a planned RRI training activity for Ph.D. students from different disciplinary areas. Via this synergetic action, QUTE4E aims to both assess its internal RRI coherence and, at the same time, provide an RRI measurement tool for education and outreach in quantum physics and technologies.

# 6 Conclusion

In this article, we introduced a wide range of quantum games and interactive tools that have been used across the community for several years for QT outreach and education. Sub-sections on these tools were written by their developers for them to describe their work in their own words, to share their experiences from the field of games/tools development for outreach and education purposes. Additionally, we covered examples of two different types of activities, quantum game hackathons/jams, and semester projects for quantum game development. Finally, we described the QUTE4E pilot project under the QTEdu Coordination and Support Action within the European Quantum Flagship Program as an example of how utilization and development activities of such tools can be integrated with quantum outreach and education efforts. These examples were provided to give readers guidance in formulating creative projects and enhance institutions' capacity to launch pioneering initiatives.

The wide range of games and tools provided here cannot be evaluated from a comparative point of view due to the lack of comprehensive assessment tools in literature. It is important to develop approaches to evaluate the performance efficiency and effectiveness of the games and tools presented here. The challenge lies in identifying the purpose and scope of their use as they address diverse needs in education, outreach, and research.

The quote "*Quantum computing is a marathon not a sprint*"[189] is widely used within the community. It is important to add here that this marathon is one that has significant implications for the future of technology and its impact on society[13,15]. Hence, widest possible participation and engagement by the stakeholders within society should be targeted. We hope that the illustrative examples provided in this article, including games, tools, events, and models, can serve as a solid departure point for current and future practitioners of quantum outreach and education.

## Acknowledgements

BL and ND would like to acknowledge the support of the Office of Naval Research (Grant No. N00014-18-1-2233) and the National Science Foundation (Grant No. 1842086) and to thank the students in the Fall 2021 Quantum Computing FRI stream.

LN would like to acknowledge the support from UKRI grant BB/T018666/1.

JW would like to acknowledge support from the Swiss National Science Foundation through NCCR SPIN.

*ScienceAtHome* would like to acknowledge grants from Carlsberg Foundation and the European Union's Horizon 2020 research and innovation programme under the Marie Skłodowska-Curie QuSCo Grant Agreement No. 765267 and the ERC PoC grant 899930.

*Quantum Flytrap* would like to acknowledge the support of the Centre for Quantum Technologies, National University of Singapore, an eNgage – III/2014 grant by the Foundation for Polish Science, and a Unitary Fund microgrant.

ZCS would like to acknowledge that their research is supported by the DAAD.

# 7 References

1. J. P. Dowling and G. J. Milburn, "Quantum technology: the second quantum revolution," 1809, Philos. Trans. R. Soc. Lond. Ser. Math. Phys. Eng. Sci. **361**(1809), 1655–1674, Royal Society (2003) [doi:10.1098/rsta.2003.1227].
2. P. Knight and I. Walmsley, "UK national quantum technology programme," 4, Quantum Sci. Technol. **4**(4), 040502, IOP Publishing (2019) [doi:10.1088/2058-9565/ab4346].
3. M. G. Raymer and C. Monroe, "The US National Quantum Initiative," 2, Quantum Sci. Technol. **4**(2), 020504, IOP Publishing (2019) [doi:10.1088/2058-9565/ab0441].
4. M. Riedel et al., "Europe's Quantum Flagship initiative," 2, Quantum Sci. Technol. **4**(2), 020501, IOP Publishing (2019) [doi:10.1088/2058-9565/ab042d].
5. "Quantum plan," Gouvernement.fr, 6 January 2022, <https://www.gouvernement.fr/en/quantum-plan> (accessed 6 January 2022).
6. É. Kelly, "Germany to invest €2B in quantum technologies," Science|Business, <https://sciencebusiness.net/news/germany-invest-eu2b-quantum-technologies> (accessed 11 February 2022).
7. T. V. Padma, "India bets big on quantum technology," Nature (2020) [doi:10.1038/d41586-020-00288-x].
8. Q. Zhang et al., "Quantum information research in China," 4, Quantum Sci. Technol. **4**(4), 040503, IOP Publishing (2019) [doi:10.1088/2058-9565/ab4bea].
9. Z. C. Seskir and A. U. Aydinoglu, "The landscape of academic literature in quantum technologies," 02, Int. J. Quantum Inf. **19**(02), 2150012, World Scientific Publishing Co. (2021) [doi:10.1142/S021974992150012X].
10. Z. C. Seskir and K. W. Willoughby, "Global innovation and competition in quantum technology, viewed through the lens of patents and artificial intelligence," Int. J. Intellect. Prop. Manag. **12**(1) (2022) [doi:10.1504/IJIPM.2021.10044326].
11. E. Gibney, "Quantum gold rush: the private funding pouring into quantum start-ups," Nature **574**(7776), 22–24 (2019) [doi:10.1038/d41586-019-02935-4].
12. S. Jarman, "UK public has little understanding of quantum technologies, says survey," Physics World, 22 June 2018, <https://physicsworld.com/a/uk-public-has-little-understanding-of-quantum-technologies-says-survey/> (accessed 25 January 2022).
13. P. E. Vermaas, "The societal impact of the emerging quantum technologies: a renewed urgency to make quantum theory understandable," 4, Ethics Inf. Technol. **19**(4), 241–246 (2017) [doi:10.1007/s10676-017-9429-1].
14. C. Hughes et al., "Assessing the Needs of the Quantum Industry," ArXiv210903601 Phys. Physicsquant-Ph (2021).
15. R. de Wolf, "The potential impact of quantum computers on society," 4, Ethics Inf. Technol. **19**(4), 271–276 (2017) [doi:10.1007/s10676-017-9439-z].
16. C. Coenen and A. Grunwald, "Responsible research and innovation (RRI) in quantum technology," Ethics Inf. Technol. **19**(4), 277–294 (2017) [doi:10.1007/s10676-017-9432-6].
17. T. Roberson, "Building a common language for interdisciplinary responsible innovation: Talking about responsible quantum in Australia," ArXiv211201378 Phys. Physicsquant-Ph (2021).
18. C. Foti et al., "Quantum Physics Literacy Aimed at K12 and the General Public," 4, Universe **7**(4), 86, Multidisciplinary Digital Publishing Institute (2021) [doi:10.3390/universe7040086].
19. L. Nita et al., "The challenge and opportunities of quantum literacy for future education and transdisciplinary problem-solving," Res. Sci. Technol. Educ. **0**(0), 1–17, Routledge (2021) [doi:10.1080/02635143.2021.1920905].
20. G. Wolbring, "Auditing the 'Social' of Quantum Technologies: A Scoping Review," 2, Societies **12**(2), 41, Multidisciplinary Digital Publishing Institute (2022) [doi:10.3390/soc12020041].

21. C. Macchiavello, "Quantum Technology - Education Coordination and Support actions," Quantum Technology, <https://qt.eu/about-quantum-flagship/projects/education-coordination-support-actions/> (accessed 25 January 2022).
22. "Home | National Q-12 Education Partnership | UIUC," <https://q12education.org/> (accessed 25 January 2022).
23. "IEEE Quantum Education," <https://ed.quantum.ieee.org/> (accessed 25 January 2022).
24. "QuanTime," <https://q12education.org/quantime> (accessed 18 April 2022).
25. F. S. Khan et al., "Quantum games: a review of the history, current state, and interpretation," Quantum Inf. Process. **17**(11), 1–42 (2018).
26. S. Phoenix, F. Khan, and B. Teklu, "Preferences in quantum games," Phys. Lett. A **384**(15), 126299 (2020) [doi:10.1016/j.physleta.2020.126299].
27. K. Schrier, *Knowledge Games: How Playing Games Can Solve Problems, Create Insight, and Make Change*, Johns Hopkins University Press, Baltimore (2016).
28. "Quantum Technologies Education for Everyone [QuTE4E]," QTEdu, <https://qtedu.eu/project/quantum-technologies-education-everyone> (accessed 29 March 2022).
29. P. Migdał, "Science-based games," 25 January 2022, <https://github.com/stared/science-based-games-list> (accessed 26 January 2022).
30. P. A. M. Dirac, *The Principles of Quantum Mechanics*, 4th edition, Clarendon Press, Oxford (1982).
31. B.-G. Englert, *Lectures on Quantum Mechanics - Volume 1: Basic Matters*, World Scientific Publishing Company, Hackensack, N.J (2006).
32. D. J. Griffiths, *Introduction to Quantum Mechanics*, 3rd edition, Cambridge University Press, Cambridge ; New York, NY (2018).
33. R. Shankar, *Principles of Quantum Mechanics, 2nd Edition*, 2nd edition, Plenum Press, New York (1994).
34. C. Roychoudhuri, "Can classical optical superposition principle get us out of quantum mysticism of non-locality and bring back reality to modern physics?," presented at Tenth International Topical Meeting on Education and Training in Optics and Photonics, 3 August 2015, Ottawa, Ontario, Canada, 966508 [doi:10.1117/12.2207464].
35. M. Beck and E. Dederick, "Quantum optics laboratories for undergraduates," in 12th Education and Training in Optics and Photonics Conference **9289**, pp. 359–366, SPIE (2014) [doi:10.1117/12.2070525].
36. E. J. Galvez, "Quantum optics laboratories for teaching quantum physics," in Fifteenth Conference on Education and Training in Optics and Photonics: ETOP 2019 **11143**, pp. 312–318, SPIE (2019) [doi:10.1117/12.2523843].
37. S. G. Lukishova, "Quantum optics and nano-optics teaching laboratory for the undergraduate curriculum: teaching quantum mechanics and nano-physics with photon counting instrumentation," in 14th Conference on Education and Training in Optics and Photonics: ETOP 2017 **10452**, pp. 508–527, SPIE (2017) [doi:10.1117/12.2269872].
38. M. T. Posner et al., "Taking large optical quantum states out of the lab: engaging pupils and the public on quantum photonics sciences," in Optics Education and Outreach V **10741**, pp. 86–91, SPIE (2018) [doi:10.1117/12.2319391].
39. M. A. Nielsen and I. L. Chuang, *Quantum Computation and Quantum Information: 10th Anniversary Edition*, 1st edition, Cambridge University Press, Cambridge ; New York (2011).
40. L. Susskind and G. Hrabovsky, *The Theoretical Minimum: What You Need to Know to Start Doing Physics*, Reprint edition, Basic Books, New York (2014).
41. V. Scarani, L. Chua, and S. Y. Liu, *Six Quantum Pieces: A First Course in Quantum Physics*, WORLD SCIENTIFIC (2010) [doi:10.1142/7965].
42. P. Migdał, "Quantum mechanics for high-school students," Piotr Migdał - blog, 15 August 2016, <https://p.migdal.pl/2016/08/15/quantum-mechanics-for-high-school-students.html> (accessed 11 February 2022).

43. D. Kaiser, *How the Hippies Saved Physics: Science, Counterculture, and the Quantum Revival*, Reprint Edition, Norton & Company, New York London (2012).
44. C. Singh, "Student understanding of quantum mechanics," Am. J. Phys. **69**(8), 885–895, American Association of Physics Teachers (2001) [doi:10.1119/1.1365404].
45. D. F. Styer, "Common misconceptions regarding quantum mechanics," Am. J. Phys. **64**(1), 31–34, American Association of Physics Teachers (1996) [doi:10.1119/1.18288].
46. S. Aaronson, "Better late than never," in Shtetl-Optimized (2011).
47. C. H. Bennett and G. Brassard, "Quantum cryptography: Public key distribution and coin tossing," Theor. Comput. Sci. **560**, 7–11 (2014) [doi:10.1016/j.tcs.2014.05.025].
48. A. K. Ekert, "Quantum Cryptography and Bell's Theorem," in Quantum Measurements in Optics, P. Tombesi and D. F. Walls, Eds., pp. 413–418, Springer US, Boston, MA (1992) [doi:10.1007/978-1-4615-3386-3_34].
49. S. J. Freedman and J. F. Clauser, "Experimental Test of Local Hidden-Variable Theories," Phys. Rev. Lett. **28**(14), 938–941, American Physical Society (1972) [doi:10.1103/PhysRevLett.28.938].
50. A. Aspect, P. Grangier, and G. Roger, "Experimental Tests of Realistic Local Theories via Bell's Theorem," Phys. Rev. Lett. **47**(7), 460–463, American Physical Society (1981) [doi:10.1103/PhysRevLett.47.460].
51. S. Aaronson and A. Arkhipov, "The Computational Complexity of Linear Optics," ArXiv10113245 Quant-Ph (2010).
52. P. Migdał, J. Rodriguez-Laguna, and M. Lewenstein, "Entanglement classes of permutation-symmetric qudit states: Symmetric operations suffice," Phys. Rev. A **88**(1), 012335, American Physical Society (2013) [doi:10.1103/PhysRevA.88.012335].
53. C. Sparrow et al., "Simulating the vibrational quantum dynamics of molecules using photonics," 7707, Nature **557**(7707), 660–667, Nature Publishing Group (2018) [doi:10.1038/s41586-018-0152-9].
54. N. D. Mermin, "From Cbits to Qbits: Teaching computer scientists quantum mechanics," Am. J. Phys. **71**(1), 23–30, American Association of Physics Teachers (2003) [doi:10.1119/1.1522741].
55. S. Aaronson, *Quantum Computing since Democritus*, Cambridge University Press, Cambridge (2013) [doi:10.1017/CBO9780511979309].
56. Ö. Salehi, Z. Seskir, and İ. Tepe, "A Computer Science-Oriented Approach to Introduce Quantum Computing to a New Audience," IEEE Trans. Educ., 1–8 (2021) [doi:10.1109/TE.2021.3078552].
57. M. Mykhailova and K. M. Svore, "Teaching Quantum Computing through a Practical Software-driven Approach: Experience Report," in Proceedings of the 51st ACM Technical Symposium on Computer Science Education, pp. 1019–1025, Association for Computing Machinery, New York, NY, USA (2020).
58. S. E. Economou, T. Rudolph, and E. Barnes, "Teaching quantum information science to high-school and early undergraduate students," ArXiv200507874 Phys. Physicsquant-Ph (2020).
59. A. Perry et al., "Quantum Computing as a High School Module," ArXiv190500282 Phys. Physicsquant-Ph (2020).
60. J. R. Wootton et al., "Teaching quantum computing with an interactive textbook," in 2021 IEEE International Conference on Quantum Computing and Engineering (QCE), pp. 385–391 (2021) [doi:10.1109/QCE52317.2021.00058].
61. A. Anupam et al., "Particle in a Box: An Experiential Environment for Learning Introductory Quantum Mechanics," IEEE Trans. Educ. **61**(1), 29–37 (2018) [doi:10.1109/TE.2017.2727442].
62. R. Peng et al., "Interactive visualizations for teaching quantum mechanics and semiconductor physics," in 2014 IEEE Frontiers in Education Conference (FIE) Proceedings, pp. 1–4 (2014) [doi:10.1109/FIE.2014.7044207].
63. M. Tople, "A Novel Interactive Paradigm for Teaching Quantum Mechanics," 2016, <https://www.semanticscholar.org/paper/A-Novel-Interactive-Paradigm-for-Teaching-Quantum-Tople/50c45da9ff05aa59e4d95f4f5e650866a22d7ee5> (accessed 26 January 2022).
64. QWorld, "Mapping the Landscape of Quantum Education," Mapping the Landscape of Quantum Education, <https://qmap.qworld.net/> (accessed 27 January 2022).

65. "Report on how to support high school teachers interested in teaching and mentoring students in quantum science | IEEE Quantum Education," <https://ed.quantum.ieee.org/teaching-high-school-discussion-report/> (accessed 27 January 2022).
66. N. Skult and J. Smed, "The Marriage of Quantum Computing and Interactive Storytelling," in Games and Narrative: Theory and Practice, B. Bostan, Ed., pp. 191–206, Springer International Publishing, Cham (2022) [doi:10.1007/978-3-030-81538-7_13].
67. J. R. Hiller, I. D. Johnston, and D. F. Styer, *Quantum Mechanics Simulations: The Consortium for Upper-Level Physics Software*, 1st edition, Wiley, New York (1995).
68. J. R. Hiller, D. F. Styer, and I. D. Johnston, *Quantum Mechanics Simulations*, John Wiley And Sons Ltd (2000).
69. B. Thaller, *Visual Quantum Mechanics: Selected Topics with Computer-Generated Animations of Quantum-Mechanical Phenomena*, Springer New York, New York, NY (2000) [doi:10.1007/b98962].
70. B. Thaller, *Advanced Visual Quantum Mechanics*, Springer-Verlag, New York (2005) [doi:10.1007/b138654].
71. D. A. Zollman, N. S. Rebello, and K. Hogg, "Quantum mechanics for everyone: Hands-on activities integrated with technology," Am. J. Phys. **70**(3), 252–259, American Association of Physics Teachers (2002) [doi:10.1119/1.1435347].
72. P. Falstad, "Math, Physics, and Engineering Applets," <http://www.falstad.com/mathphysics.html> (accessed 11 February 2022).
73. W. Christian et al., "The Physlet Approach to Simulation Design," Phys. Teach. **53**(7), 419–422, American Association of Physics Teachers (2015) [doi:10.1119/1.4931011].
74. M. Joffre, "Quantum Physics Online," <https://www.quantum-physics.polytechnique.fr/> (accessed 11 February 2022).
75. "PhET Interactive Simulations," PhET, <https://phet.colorado.edu/> (accessed 11 February 2022).
76. C. E. Wieman et al., "Teaching Physics Using PhET Simulations," Phys. Teach. **48**(4), 225–227, American Association of Physics Teachers (2010) [doi:10.1119/1.3361987].
77. W. K. Adams et al., "A Study of Educational Simulations Part I - Engagement and Learning.," J. Interact. Learn. Res. **19**(3), 397–419, Association for the Advancement of Computing in Education (AACE) (2008).
78. W. K. Adams et al., "A Study of Educational Simulations Part II – Interface Design," J. Interact. Learn. Res. **19**(4), 551–577, Association for the Advancement of Computing in Education (AACE) (2008).
79. N. S. Podolefsky et al., "Computer simulations to classrooms: tools for change," presented at 2009 PHYSICS EDUCATION RESEARCH CONFERENCE, 2009, Ann Arbor (MI), 233–236 [doi:10.1063/1.3266723].
80. S. Chandralekha, "QuILT homepage," PhysPort, <https://www.physport.org/curricula/quilts/> (accessed 11 February 2022).
81. S. DeVore, E. Marshman, and C. Singh, "Challenge of engaging all students via self-paced interactive electronic learning tutorials for introductory physics," Phys. Rev. Phys. Educ. Res. **13**(1), 010127, American Physical Society (2017) [doi:10.1103/PhysRevPhysEducRes.13.010127].
82. "QuVis: The Quantum Mechanics Visualisation Project," <https://www.st-andrews.ac.uk/physics/quvis/> (accessed 11 February 2022).
83. A. Kohnle et al., "A new multimedia resource for teaching quantum mechanics concepts," Am. J. Phys. **80**(2), 148–153, American Association of Physics Teachers (2012) [doi:10.1119/1.3657800].
84. A. Kohnle et al., "Developing and evaluating animations for teaching quantum mechanics concepts," Eur. J. Phys. **31**(6), 1441–1455, IOP Publishing (2010) [doi:10.1088/0143-0807/31/6/010].
85. "Quantum Composer," <https://www.quatomic.com/composer/> (accessed 11 February 2022).
86. S. Zaman Ahmed et al., "Quantum composer: A programmable quantum visualization and simulation tool for education and research," Am. J. Phys. **89**(3), 307–316, American Association of Physics Teachers (2021) [doi:10.1119/10.0003396].

87. C. A. Weidner et al., "Investigating student use of a flexible tool for simulating and visualizing quantum mechanics," presented at 2020 Physics Education Research Conference Proceedings, 1 September 2020, 563–568.
88. J. Schell, "Art of Game Design," Schell Games, <https://www.schellgames.com/art-of-game-design/> (accessed 11 February 2022).
89. E. Bonawitz et al., "The double-edged sword of pedagogy: Instruction limits spontaneous exploration and discovery," Cognition **120**(3), 322–330 (2011) [doi:10.1016/j.cognition.2010.10.001].
90. "A Slower Speed of Light – MIT Game Lab."
91. "The Electric Shocktopus," <http://testtubegames.com/shocktopus.html> (accessed 11 February 2022).
92. "Kerbal Space Program – Create and Manage Your Own Space Program."
93. O. J. Kjær, "NandGame - Build a computer from scratch.," <https://nandgame.com/about> (accessed 11 February 2022).
94. "Games sales," SteamSpy - All the data about Steam games, <https://steamspy.com/> (accessed 11 February 2022).
95. N. Case, "Explorable Explanations," 4 January 2022, <https://github.com/explorableexplanations/explorableexplanations.github.io> (accessed 11 February 2022).
96. B. Victor, "Explorable Explanations," <http://worrydream.com/ExplorableExplanations/> (accessed 11 February 2022).
97. A. Matuschak and M. Nielsen, "How can we develop transformative tools for thought?" (2019).
98. M. Adereth, "Colorful Equations with MathJax - adereth," 29 November 2013, <https://adereth.github.io/blog/2013/11/29/colorful-equations/> (accessed 15 February 2022).
99. S. Riffle, "Understanding the Fourier transform," 2011, <https://web.archive.org/web/20130318211259/http://www.altdevblogaday.com/2011/05/17/understanding-the-fourier-transform> (accessed 15 February 2022).
100. O. Byrne, *The First Six Books of the Elements of Euclid, in Which Coloured Diagrams and Symbols are Used Instead of Letters for the Greater Ease of learners*, Isha Books (2013).
101. P. Migdał, B. Olechno, and B. Podgórski, "Level generation and style enhancement -- deep learning for game development overview," ArXiv210707397 Cs (2021).
102. P. Migdał, "Interactive Machine Learning List," <https://p.migdal.pl/interactive-machine-learning-list/> (accessed 11 February 2022).
103. D. Kunin, "Seeing Theory," <http://seeingtheory.io> (accessed 11 February 2022).
104. S. Yee and T. Chu, "A visual introduction to machine learning," 10 February 2022, <http://www.r2d3.us/visual-intro-to-machine-learning-part-1/> (accessed 11 February 2022).
105. D. Smilkov and S. Carter, "Tensorflow — Neural Network Playground," <http://playground.tensorflow.org> (accessed 11 February 2022).
106. C. Olah, A. Mordvintsev, and L. Schubert, "Feature Visualization," Distill **2**(11), 10.23915/distill.00007 (2017) [doi:10.23915/distill.00007].
107. C. Olah et al., "The Building Blocks of Interpretability," Distill **3**(3), e10 (2018) [doi:10.23915/distill.00010].
108. "Distill Hiatus," Distill **6**(7), e31 (2021) [doi:10.23915/distill.00031].
109. B. M. Randles et al., "Using the Jupyter Notebook as a Tool for Open Science: An Empirical Study," in 2017 ACM/IEEE Joint Conference on Digital Libraries (JCDL), pp. 1–2 (2017) [doi:10.1109/JCDL.2017.7991618].
110. J. M. Perkel, "Why Jupyter is data scientists' computational notebook of choice," Nature **563**(7729), 145–146 (2018) [doi:10.1038/d41586-018-07196-1].
111. A. Matuschak and M. Nielsen, "Quantum Country," 2019, <https://quantum.country> (accessed 11 February 2022).

112. C. Gidney, "Building your own Quantum Fourier Transform," <https://algassert.com/quantum/2014/03/07/Building-your-own-Quantum-Fourier-Transform.html> (accessed 11 February 2022).
113. C. Zendejas-Morales and P. Migdał, "Quantum logic gates for a single qubit, interactively," <https://quantumflytrap.com/blog/2021/qubit-interactively/> (accessed 11 February 2022).
114. C. Cantwell, "Quantum Chess: Developing a Mathematical Framework and Design Methodology for Creating Quantum Games," ArXiv190605836 Quant-Ph (2019).
115. A. Sharma, "Mapping the Landscape of Quantum Games."
116. J. H. M. Jensen et al., "Crowdsourcing human common sense for quantum control," Phys. Rev. Res. **3**(1), 013057, American Physical Society (2021) [doi:10.1103/PhysRevResearch.3.013057].
117. L. Nita et al., "Inclusive learning for quantum computing: supporting the aims of quantum literacy using the puzzle game Quantum Odyssey," ArXiv210607077 Phys. Physicsquant-Ph (2021).
118. "PhysPort: Supporting physics teaching with research-based resources," PhysPort, <https://www.physport.org/> (accessed 29 March 2022).
119. D. Hestenes, M. Wells, and G. Swackhamer, "Force concept inventory," Phys. Teach. **30**(3), 141–158, American Association of Physics Teachers (1992) [doi:10.1119/1.2343497].
120. M. Fingerhuth, T. Babej, and P. Wittek, "Open source software in quantum computing," PLOS ONE **13**(12), e0208561, Public Library of Science (2018) [doi:10.1371/journal.pone.0208561].
121. M. L. Chiofalo et al., "Games for Quantum Physics Education," presented at Symposium at WCPE 2021, 2022, (submitted).
122. S. Goorney et al., "Culturo-scientific Storytelling," 2022, (submitted).
123. M. L. Chiofalo et al., "Games for Teaching/Learning Quantum Mechanics: a pilot study with high-school students," 2022, (in preparation).
124. A. Drachen, P. Mirza-Babaei, and L. Nacke, Eds., *Games User Research*, Oxford University Press, Oxford, UK ; New York (2018).
125. C. C. Abt, *Serious Games*, University Press of America, Lanham, MD (2002).
126. T. M. Connolly et al., "A systematic literature review of empirical evidence on computer games and serious games," Comput. Educ. **59**(2), 661–686 (2012) [doi:10.1016/j.compedu.2012.03.004].
127. P. Milgram et al., "Augmented reality: a class of displays on the reality-virtuality continuum," in Telemanipulator and Telepresence Technologies **2351**, pp. 282–292, SPIE (1995) [doi:10.1117/12.197321].
128. "IBM Quantum," IBM Quantum, <https://quantum-computing.ibm.com/> (accessed 11 February 2022).
129. "Hello Quantum," <https://helloquantum.mybluemix.net/> (accessed 26 January 2022).
130. "Learn Quantum Computation using Qiskit," <https://community.qiskit.org/textbook/preface.html> (accessed 11 February 2022).
131. J. R. Wootton, "Visualizing bits and qubits," Qiskit, 18 May 2018, <https://medium.com/qiskit/visualizing-bits-and-qubits-9af287047b28> (accessed 11 February 2022).
132. J. R. Wootton, "Getting started with the IBM Q Experience," Qiskit, 29 March 2021, <https://medium.com/qiskit/how-to-turn-on-your-quantum-computer-fba0a4152d92> (accessed 11 February 2022).
133. J. R. Wootton, "How to program a quantum computer," Qiskit, 10 December 2020, <https://medium.com/qiskit/how-to-program-a-quantum-computer-982a9329ed02> (accessed 11 February 2022).
134. A. Anupam et al., "Design Challenges for Science Games:: The Case of a Quantum Mechanics Game," 1, Int. J. Des. Learn. **11**(1), 1–20 (2020) [doi:10.14434/ijdl.v11i1.24264].
135. A. Anupam et al., "Beyond Motivation and Memorization: Fostering Scientific Inquiry with Games," in Extended Abstracts of the Annual Symposium on Computer-Human Interaction in Play Companion Extended Abstracts, pp. 323–331, Association for Computing Machinery, New York, NY, USA (2019) [doi:10.1145/3341215.3356309].
136. "QPlayLearn – Q from A to Z," <https://qplaylearn.com/> (accessed 26 January 2022).

137. H. E. Gardner, *Frames of Mind: The Theory of Multiple Intelligences*, 3rd edition, Basic Books, New York (2011).
138. P. Migdał et al., "Visualizing quantum mechanics in an interactive simulation - Virtual Lab by Quantum Flytrap," ArXiv220313300 Quant-Ph (2022).
139. M. H. Göker and H. Birkhofer, "Problem Solving with 'The Incredible Machine' - An Experiment in Case-Based Reasoning," in Proceedings of the First International Conference on Case-Based Reasoning Research and Development, pp. 441–450, Springer-Verlag, Berlin, Heidelberg (1995).
140. P. Migdał, P. Hes, and M. Krupiński, "Quantum Game with Photons (2014-2016)," 20 January 2022, <https://github.com/stared/quantum-game> (accessed 11 February 2022).
141. B. Dorland et al., "Quantum Physics vs. Classical Physics: Introducing the Basics with a Virtual Reality Game," in Games and Learning Alliance, A. Liapis et al., Eds., pp. 383–393, Springer International Publishing, Cham (2019) [doi:10.1007/978-3-030-34350-7_37].
142. S. Ornes, "Science and Culture: Quantum games aim to demystify heady science," Proc. Natl. Acad. Sci. **115**(8), 1667–1669, National Academy of Sciences (2018) [doi:10.1073/pnas.1800744115].
143. A. Parakh et al., "A Novel Approach for Embedding and Traversing Problems in Serious Games," in Proceedings of the 21st Annual Conference on Information Technology Education, pp. 229–235, Association for Computing Machinery, New York, NY, USA (2020) [doi:10.1145/3368308.3415417].
144. A. Parakh, P. Chundi, and M. Subramaniam, "An Approach Towards Designing Problem Networks in Serious Games," in 2019 IEEE Conference on Games (CoG), pp. 1–8 (2019) [doi:10.1109/CIG.2019.8848055].
145. J. D. Weisz, M. Ashoori, and Z. Ashktorab, "Entanglion: A Board Game for Teaching the Principles of Quantum Computing," in Proceedings of the 2018 Annual Symposium on Computer-Human Interaction in Play, pp. 523–534, Association for Computing Machinery, New York, NY, USA (2018) [doi:10.1145/3242671.3242696].
146. Z. Ashktorab, J. D. Weisz, and M. Ashoori, "Thinking Too Classically: Research Topics in Human-Quantum Computer Interaction," in Proceedings of the 2019 CHI Conference on Human Factors in Computing Systems, pp. 1–12, Association for Computing Machinery, New York, NY, USA (2019) [doi:10.1145/3290605.3300486].
147. M. Leifer, "Gamifying Quantum Theory," Math. Phys. Comput. Sci. Fac. Artic. Res. (2017).
148. P. Migdał and P. Cochin, "*Quantum Tensors - an NPM package for sparse matrix operations for quantum information and computing*", <https://github.com/Quantum-Flytrap/quantum-tensors> (2022).
149. K. Jankiewicz and P. Migdał, "BraKetVue - a Vue-based visualization of quantum states and operations*"*, <https://github.com/Quantum-Flytrap/bra-ket-vue> (2022).
150. "Quarks Interactive," Quarks Interactive, <https://www.quarksinteractive.com/> (accessed 11 February 2022).
151. "ScienceAtHome | Citizen science games," ScienceAtHome.org, <https://www.scienceathome.org/> (accessed 26 January 2022).
152. "ScienceAtHome | Games | Quantum Moves 2," ScienceAtHome.org, <https://www.scienceathome.org/games/quantum-moves-2/> (accessed 11 February 2022).
153. "The Alice Framework," ScienceAtHome.org, <https://www.scienceathome.org/games/the-alice-framework/> (accessed 11 February 2022).
154. M. H. Anderson et al., "Observation of Bose-Einstein Condensation in a Dilute Atomic Vapor," Science **269**(5221), 198–201, American Association for the Advancement of Science (1995) [doi:10.1126/science.269.5221.198].
155. K. B. Davis et al., "Bose-Einstein Condensation in a Gas of Sodium Atoms," Phys. Rev. Lett. **75**(22), 3969–3973, American Physical Society (1995) [doi:10.1103/PhysRevLett.75.3969].
156. J. S. Laustsen et al., "Remote multi-user control of the production of Bose–Einstein condensates," Appl. Phys. B **127**(9), 125 (2021) [doi:10.1007/s00340-021-07671-0].

157. R. Heck et al., "Remote optimization of an ultracold atoms experiment by experts and citizen scientists," Proc. Natl. Acad. Sci. **115**(48), E11231–E11237, National Academy of Sciences (2018) [doi:10.1073/pnas.1716869115].
158. "QEngine," <https://www.quatomic.com/qengine/> (accessed 11 February 2022).
159. J. J. Sørensen et al., "QEngine: A C++ library for quantum optimal control of ultracold atoms," Comput. Phys. Commun. **243**, 135–150 (2019) [doi:10.1016/j.cpc.2019.04.020].
160. "ReGAME Cup18," ScienceAtHome.org, <https://www.scienceathome.org/education/regame-cup-2018/> (accessed 11 February 2022).
161. "Forskningsspil i Folkeskolen," ScienceAtHome.org, <https://www.scienceathome.org/education/fif/> (accessed 11 February 2022).
162. "QuSCo | Quantum-enhanced Sensing via Quantum Control," <https://qusco-itn.eu/> (accessed 11 February 2022).
163. S. Z. Ahmed et al., "A training programme for early-stage researchers that focuses on developing personal science outreach portfolios," ArXiv210303109 Phys. Physicsquant-Ph (2021).
164. B. R. La Cour et al., "The Virtual Quantum Optics Laboratory," ArXiv210507300 Quant-Ph (2021).
165. "Virtual Quantum Optics Lab," <https://www.vqol.org/> (accessed 11 February 2022).
166. B. R. La Cour and T. W. Yudichak, "Classical model of a delayed-choice quantum eraser," Phys. Rev. A **103**(6), 062213, American Physical Society (2021) [doi:10.1103/PhysRevA.103.062213].
167. J. A. Walsh et al., "Piloting a full-year, optics-based high school course on quantum computing," Phys. Educ. **57**(2), 025010, IOP Publishing (2021) [doi:10.1088/1361-6552/ac3dc2].
168. "OpenBSD: Hackathons," <http://www.openbsd.org/hackathons.html> (accessed 11 February 2022).
169. "Definition of hackathon | Dictionary.com," www.dictionary.com, <https://www.dictionary.com/browse/hackathon> (accessed 11 February 2022).
170. C. Hecker, "Indie Game Jam 0," <http://www.indiegamejam.com/igj0/> (accessed 11 February 2022).
171. "The complete guide to organizing a successful hackathon | HackerEarth," in Innovation Management Resources (2017).
172. "2014 Games – Quantum Game Jam," <http://www.finnishgamejam.com/quantumjam2015/games/2014-games/> (accessed 26 January 2022).
173. J. R. Wootton, "The History of Games for Quantum Computers," Medium, 6 April 2020, <https://decodoku.medium.com/the-history-of-games-for-quantum-computers-a1de98859b5a> (accessed 26 January 2022).
174. A. Kultima, L. Piispanen, and M. Junnila, "Quantum Game Jam - Making Games with Quantum Physicists," in Academic Mindtrek 2021, pp. 134–144, Association for Computing Machinery, New York, NY, USA (2021).
175. "Quantum AI Foundation - Quantum Games Hackathon," <https://www.qaif.org/contests/quantum-games-hackathon> (accessed 26 January 2022).
176. "Quantum AI Foundation," <https://www.qaif.org/> (accessed 11 February 2022).
177. "QPoland – QWorld," <https://qworld.net/qpoland/> (accessed 11 February 2022).
178. "QWorld – Be part of the second quantum revolution!," <https://qworld.net/> (accessed 11 February 2022).
179. "Quantum AI Foundation - Warsaw Quantum Computing Group," <https://www.qaif.org/events/warsaw-quantum-computing-group> (accessed 11 February 2022).
180. "QBronze – QWorld," <https://qworld.net/workshop-bronze/> (accessed 11 February 2022).
181. "Quantum AI Foundation - YouTube," <https://www.youtube.com/channel/UCoQAyPU5KQEpMOMDUN0j3IQ/videos> (accessed 11 February 2022).

182. B. Coecke, *Picturing Quantum Processes: A First Course in Quantum Theory and Diagrammatic Reasoning*, 1st edition, Cambridge University Press, Cambridge, United Kingdom ; New York, NY, USA (2017).
183. "Labster | 200+ virtual labs for universities and high schools," <https://www.labster.com/> (accessed 11 February 2022).
184. M. Pavlidou and C. Lazzeroni, "Particle physics for primary schools—enthusing future physicists," 5, Phys. Educ. **51**(5), 054003, IOP Publishing (2016) [doi:10.1088/0031-9120/51/5/054003].
185. C. Lazzeroni, S. Malvezzi, and A. Quadri, "Teaching Science in Today's Society: The Case of Particle Physics for Primary Schools," 6, Universe **7**(6), 169, Multidisciplinary Digital Publishing Institute (2021) [doi:10.3390/universe7060169].
186. "Quantum Tour," <https://quantumtour.icfo.eu/> (accessed 11 February 2022).
187. Robin Baumgarten, *Quantum Jungle: Interactive Quantum Art Installation with 1008 springs!* (2021).
188. "Quantum Jungle – QPlayLearn," <https://qplaylearn.com/quantum-jungle> (accessed 11 February 2022).
189. C. Monroe, "Quantum computing is a marathon not a sprint," in VentureBeat (2019).